\documentclass[aps,onecolumn,notitlepage,floatfix,showpacs,citeautoscript,longbibliography,hyperlinks]{revtex4-1}

\usepackage{graphicx}
\usepackage{dcolumn}
\usepackage{bm}
\usepackage{bbold}
\usepackage{mathtools}
\usepackage{color}
\usepackage{transparent}
\usepackage[colorlinks=true,citecolor=blue]{hyperref}

\usepackage{braket}
\usepackage{dsfont}

\begin{document}

\title{Geometric Entanglement and Quantum Phase Transition\\in Generalized Cluster-XY models}

\author{Aydin Deger$^{1,2}$}
\author{Tzu-Chieh Wei$^{1}$}
\affiliation{
$^1$C.N. Yang Institute for Theoretical Physics and Department of Physics and Astronomy, State University of New York at Stony Brook, Stony Brook, NY 11794-3840, USA\\$^2$Department of Applied Physics, Aalto University, 00076 Aalto, Finland
}
	
	\begin{abstract}
In this work, we investigate quantum phase transition (QPT) in a generic family of spin chains using the ground-state energy, the energy gap, and the geometric measure of entanglement (GE). In many of prior works, GE per site was used. Here, we also consider GE per block with each block size being two. This can be regarded as a coarse grain of GE per site. We introduce a useful parameterization for the family of spin chains that includes the XY models with $n$-site interaction, the GHZ-cluster model and a cluster-antiferromagnetic model, the last of which exhibits QPT between a symmetry-protected topological (SPT) phase and a symmetry-breaking antiferromagnetic phase. As the models are exactly solvable, their ground-state wavefunctions can be obtained and thus their GE can be studied. It turns out that the overlap of the ground states with translationally invariant product states can be exactly calculated and hence the GE can be obtained via further parameter optimization. The QPTs exhibited in these models are detected by the energy gap and singular behavior of geometric entanglement. In particular, the XzY model exhibits transitions from the nontrivial SPT phase to a trivial paramagnetic phase. Moreover, the halfway XY model exhibits a first-order transition across the Barouch-McCoy circle, on which it was only a crossover in the standard XY model. 
		
	\end{abstract}
	
	\maketitle
	
\section{Introduction}
	
	Quantum entanglement has now been recognized as one of many intriguing consequences of quantum physics. Nonetheless, Einstein was greatly troubled by this phenomenon. He described it as ”spooky-action-at-a-distance" in his famous EPR paper, since it seemed to imply a violation of relativistic causality \cite{Einstein1935}. Later, John Bell introduced inequalities that helped to gain more insight about quantum correlations~\cite{Bell1964} and motivated subsequent theoretical and experimental development~\cite{Clauser1969,Freedman1972,Aspect1981}. These quantum correlations have since been verified many times in different experiments \cite{Freedman1972,Aspect1981,Tittel1998,Weihs1998}. Quantum entanglement has also been found to provide resources for quantum information processing~\cite{Nielsen2000,Bennett1993,Shor1994,Grover1996,Ekert1991} and has been increasingly used as a tool for investigating in a wide range of physics from quantum computation to black holes~\cite{Maldacena2013,Steinhauer2016}. Over the years, many approaches have been devised to quantify entanglement in both bipartite and multipartite systems \cite{Amico2008,Peschel2012,Vedral1997}. The entanglement entropy is perhaps the most well known example that measures quantum correlations between two halves of a pure quantum system~\cite{Amico2008}. As another example, concurrence or the related entanglement of formation quantifies entanglement between two qubits, and, among many useful features, there is an analytic formula for that~\cite{Wootters1998}. For multipartite systems, there are various definitions but most of them are not easy to calculate~\cite{Amico2008}. Thus in this paper, we will follow some prior works and adopt a particular simple multipartite measure---the geometric measure of entanglement (GE)---to quantify entanglement for pure quantum systems and examine how it detects the quantum phase transitions (QPT) for spin systems \cite{Barnum2001,Wei2003,Wei2005}.
	
Phase transition is a phenomenon that describes a change in the state of matter as an external control parameter such as temperature or pressure is varied. Boiling of water or water freezing to ice is a temperature-driven phase transition that we experience in our daily lives. On the other hand, quantum phase transitions \cite{Sachdev2011} occur at zero temperature, and, qualitatively speaking, involve either level crossing or closing of an energy gap (between the ground and excited states) as the system size increases~\cite{Sachdev2011}. In the latter case, there is a diverging correlation length at the quantum critical point. The ground-state wavefunction is expected to exhibit a singular behavior, which can be characterized by the changes of entanglement near the critical point \cite{Osborne2002,Amico2006}. Therefore, quantum entanglement may be an alternative way to detect a quantum phase transition \cite{Osterloh2002,Amico2008}, other than the thermodynamic quantities. Besides interest from quantum information and QPT \cite{Eisler2017,Bayat2017,Laflorencie2016,Blanc2017,Ye2018,DeChiara2012}, quantum entanglement is not only a powerful theoretical concept but also has been measured in several recent experiments \cite{Zhang2011,Islam2015,Kaufman2016,Sahling2015}. In particular, (cluster) spin models can be implemented in experiments and simulated in quantum information processors \cite{Verstraete2009,Schmoll2017,Cervera-Lierta2018,Choo2018,Harris2018,Britton2012,Labuhn2016,Islam2011,Zhang2017}.
	
	Since we are interested in systems at $T=0$, we will be concerned with pure quantum many-body states, $|\Psi\rangle$ of $N$ spins, expressed in some local basis, as
	\begin{equation}
	\ket{\Psi} = \sum_{p_{1} \ldots p_{N}} \Psi_{p_{1} p_{2} \ldots p_{N}} \ket{e_{p_{1}}^{(1)} e_{p_{2}}^{(2)} \ldots e_{p_{N}}^{(N)}}.
	\end{equation}
	A simple idea to quantify its quantum correlation is to see how close $|\Psi\rangle$ can be approximated by the set of uncorrelated product states $\ket{\Phi}= \mathop{\bigotimes}_i \ket{\phi^{[i]}}$ , and thus the maximal overlap $\Lambda_{\max}(\Psi)\equiv \max_{\phi's} |\braket{\Phi|\Psi}|$ is a quantity that measures such $\ket{\Psi}$'s closeness to product states. We can choose to use the formula $E_G(\Psi)\equiv -2\log \Lambda_{\max}(\Psi)$, which we call the geometric entanglement \cite{Barnum2001,Wei2005}, to quantify the quantum correlations in the state $\ket{\Psi}$. Moreover, in choosing different forms of product states $\ket{\Phi}$, one can probe different coarse-grained levels of entanglement, and these represent different hierarchies of quantum correlations:
	\begin{align*}
	\ket{\Phi_{1}}&= \bigotimes_{i}^{N} \ket{\phi^{[i]}} \Rightarrow \ \text{entanglement among all sites},\\
	\ket{\Phi_{2}}&= \bigotimes_{i}^{N/2} \ket{\phi^{[2i-1, 2i]}} \Rightarrow \ \text{entanglement among all blocks with 2 sites},\\
	\ket{\Phi_{L}}&= \bigotimes_{i}^{N/L} \ket{\phi^{[L i-L-1,.. L i]}} \Rightarrow \ \text{entanglement among all blocks of L sites}.
	\end{align*}
	In conforming with the intuitive picture of renormalization group (RG) on states (see e.g., Ref.~\cite{Verstraete2005}), we denote $|\Psi'\rangle$ as the quantum state of $|\Psi\rangle$ after one-step of RG via merging two sites into one, and the entanglement under such a RG procedure should therefore be
	defined as follows:
	\begin{equation}
	E\left({\rm RG}(\Psi)\right)=E(\{\Psi'\})= \min_U E_1(\Psi'),
	\end{equation}
	where the unitary $U$ is of the form $U[12]\otimes U[34]\otimes\cdots\otimes
	U[2k-1,2k]\otimes\cdots$ and $\ket{\Psi'}=U\ket{\Psi}$ denotes the unitary transformation that describes the merging (therefore acts on two neighboring sites in the original lattice). But since maximization over two-site unitary $U[12]$ is equivalent to maximization over two-site state $|\phi^{[12]}\rangle$, we have that
	\begin{equation}
	\max_{\Phi_1}|\langle \Phi_1|\Psi'\rangle|=\max_{\Phi_2}|\langle \Phi_2|\Psi\rangle|,
	\end{equation}
	and thus, we see that the geometric entanglement w.r.t. product of $L$-site states is the entanglement of RG after $\log_2 L$ steps on the quantum state~\cite{Wei2010}. However, to calculate different hierarchies of entanglement is generally difficult. But as we see below, the first two in the above, equivalently, the entanglement per site and per block of two, can be calculated for a wide class of exactly solvable spin chains.
	
	The purpose of this paper is threefold. Firstly, we describe and review the procedure for diagonalizing a large family of solvable spin chains which include the XY models with $n$-site interaction, the GHZ-cluster model and a cluster-antiferromagnetic model, the last of which exhibits QPT between a symmetry-protected topological (SPT) phase and an antiferromagnetic phase (AFM). We provide a convenient parameterization of these and others, forming the family which we call the generalized cluster-XY models. In diagonalizing the Hamiltonians for finite sizes, we find and illustrate subtle points in getting the true ground state and the energy gap. Secondly, we show how to compute the geometric entanglement per site and per block of 2 sites for such systems and examine QPT on the phase diagram. As explained above, this corresponds to the first two steps in the quantum-state RG procedure. One new ingredient here is the calculation of block entanglement per two sites. Thirdly, we hope that the various examples we include will be of use to readers interested in studying QPT from the perspective of entanglement. We calculate both the energy gap and the entanglement for ground state, and use both of them for characterization of quantum phase transitions (if they exist) in various \textit{cluster-XY models}. We shall see that the family of the models is interesting and displays many peculiar properties, as discussed below. Some of the models have been studied before in terms of entanglement, such as the standard XY model, the GHZ-cluster model by Wolf et al.~\cite{Wolf2006}, and the SPT-antiferromagnetic model by Son et al.~\cite{Son2011}. One new feature is that the three-site XY model (i.e., the XzY model) exhibits a transition from $Z_2\times Z_2$ SPT phase to a paramagnetic phase \cite{Montes2012}. The general $n$-site XY (with $n$ odd) is expected to have such an SPT to paramagnetic transition~\cite{Lahtinen2015}. Moreover, among the family of the models, in the halfway XY model we find a first-order transition across the Barouch-McCoy circle, on which it was only a crossover for the standard XY model. 
	
	The structure of this paper is as follows: In Sect. II, we introduce a parameterization of the generalized Hamiltonian for the cluster-XY model with $n$-site Z mediated XX and YY interaction. With this solution, one can diagonalize many bilinear Hamiltonians by substituting related parameters, quantify entanglement and detect quantum phase transition on the phase diagram. Then we give an illustrative example of Hamiltonian for XY model with $n$-site interaction using our parameterization. In Sect. III, we introduce the geometric measure of entanglement per site and block for the multipartite systems. We quantify global entanglement by calculating the overlap of ground-state wavefunctions and certain types of product states. The resultant entanglement will be used to examine quantum phase transitions in the family of the cluster-XY models. In Sect. IV, we study several examples such as XY model with three-site interaction and \textit{halfway interaction}, whose geometric entanglement has not been analyzed before. The three-site interaction XzY model exhibits transitions from nontrivial SPT phase to a trivial paramagnetic phase. Moreover, the halfway XY model exhibits a first-order transition across the Barouch-McCoy arc, on which it is only a crossover in the standard XY model. However, the halfway Ising model has no such transition. Moreover, we present solutions of paramagnetic-ferromagnetic, GHZ-Cluster \cite{Wolf2006,Wei2010}, and symmetry-protected topological (SPT)-antiferromagnetic \cite{Son2011} transitions by using this method. We make some concluding remarks in Sect. V.
	
	\section{Parameterization of Cluster-XY Models with n-site interaction}
	
	The quantum XY model was solved by Lieb, Schultz, Mattis in 1961 \cite{Lieb1961} and later all the statistical properties were examined by many other authors \cite{McCoy1968,Barouch1970,Verstraete2004a,Pachos2004,Lou2004,Derzhko2011,Titvinidze2003,DePasquale2009}. One convenient way to investigate spin chain problems is to use either bosonic or fermionic language \cite{Auerbach2012}. For example, one can analyze the Hamiltonian by using the Holstein-Primakoff transformation \cite{Primakoff1939} for mapping spin operators to bosonic annihilation and creation operators. One can, for example, use it to study spin-wave theory in the model. For some spin-chain models, on the other hand, the fermionic approach, combining the Jordan-Wigner \cite{Lieb1961} and Bogoliubov transformations~\cite{Bardeen1957}, provides a feasible way to diagonalize the Hamiltonians that are intrinsically free fermions.
	
	The motivation of this section is to generalize one-dimensional bilinear Hamiltonians with XY interaction by introducing a systematic parameterization that describes a large family of quantum spin models, for which the ground state and its geometric entanglement are exactly solved. Similar models were discussed by Suzuki~\cite{Suzuki1971a}. Here we offer a facile parameterization that includes further bilinear Hamiltonians. In particular, we introduce a few sets of parameters to describe the Hamiltonians, and diagonalize them to determine the energy spectrum. We adopt well-established methods and discuss the subtleties of determining the ground state and the energy gap for finite systems.

	\subsection{Parameterization of Hamiltonians and their diagonalization}
	\label{sec:par}
	
	We begin by defining the Hamiltonian for which there are a few types of parameters. We only consider translational invariance and models that are exactly solvable. The parameters $N^{(x)}$ and $N^{(y)}$ are the number of $X$ and $Y$ types of blocks in the Hamiltonian, respectively, which represent $X$ or $Y$ interaction mediated by $Z$: $X\underbrace{Z...Z}_{n^{(x)}}X$ or $Y\underbrace{Z...Z}_{n^{(y)}}Y$. We have indicated the numbers of consecutive $Z$ sites for each block, $n_{l}^{(x)}$ and $n_{l'}^{(y)}$, respectively. The subscript $l$ ranges from $1$ to $N^{(x)}$ and $l'$ from $1$ to $N^{(y)}$ and they indicate different ranges in the XX and YY interactions, respectively. For example, one can build a Hamiltonian with three (e.g., $N^{(x)}=3$) $X$ interaction blocks, such as $XX$, $XZX$ and $XZZX$, and only one $Y$-type block ($N^{(y)}=1$), such as $YZZZY$. To indicate the strength of each block separately, we use parameters $J_{l}^{(x)}$ and $J_{l'}^{(y)}$. For the above example, there are four such parameters, $J_1^x$, $J_2^x$, $J_3^x$ and $J_1^y$. Finally, the parameter $h$ is the strength of the transverse field. Thus, the parameterized Hamiltonian reads:
	\begin{align}
	\label{eq:PXY}
	H_{PXY} = - \sum_{j=1}^{N} \left( \sum_{l=1}^{N^{(x)}} J_{l}^{(x)} \sigma_{j-1}^{x} \sigma^{z}_{j} \ldots \sigma^{z}_{j+n_{l}^{(x)}-1} \sigma_{j+n_{l}^{(x)}}^{x} + \sum_{l'=1}^{N^{(y)}} J_{l'}^{(y)} \sigma_{j-1}^{y} \sigma^{z}_{j} \ldots \sigma^{z}_{j+n_{l'}^{(y)}-1} \sigma_{j+n_{l'}^{(y)}}^{y} + h \sigma_{j}^{z} \right),
	\end{align}	
where $\sigma$'s are the Pauli matrices associated with spin-1/2 angular momentum operators:
	\begin{equation*}
	\sigma_{j}^{x}=\Bigg(
	\begin{matrix}
	0&1\\
	1&0
	\end{matrix}\Bigg), \quad \sigma_{j}^{y}=\Bigg(\begin{matrix}
	0&-i\\
	i&0
	\end{matrix}\Bigg), \quad \sigma_{j}^{z}=\Bigg(\begin{matrix}
	1&0\\
	0&-1
	\end{matrix}\Bigg),
	\end{equation*}
and $N$ indicates the system size. We remark that the family of models in this parameterization includes many interesting ones, {such as XY model with $n$-site interaction, the GHZ-cluster model, and the SPT-AFM models and other interesting ones that have been explored from different perspectives~\cite{Wolf2006,Son2011,Lahtinen2015}. We discuss and analyze some of these in the following.}
	
	Next, we employ the Jordan-Wigner transformation, which realizes a spin to fermion $c$ mapping:
	\begin{subequations}
		\begin{align}
		\sigma^{x}_{i}&=\prod^{i-1}_{j=1} \left( 1-2 c^{\dagger}_{j} c_{j} \right) \left( c_{i} + c^{\dagger}_{i} \right),\\
		\sigma^{y}_{i}&=-i \ \prod^{i-1}_{j=1} \left( 1-2 c^{\dagger}_{j} c_{j} \right) \left( c_{i} - c^{\dagger}_{i} \right),\\
		\sigma^{z}_{i}&=1-2 c^{\dagger}_{i} c_{i},
		\end{align}
	\end{subequations}
	where the fermionic creation and annihilation operators satisfy the canonical fermionic commutation relations $\{c_{i},c_{j}^{\dagger}\}=\delta_{ij}$.
	To impose the periodic boundary conditions for spins, we rewrite the expression $\sigma^{x}_{N} \sigma^{x}_{N+1} = \sigma^{x}_{N} \sigma^{x}_{1}$ as fermions:
	\begin{subequations}
		\begin{align}
		& \left( c_{N} + c^{\dagger}_{N} \right) \left( c_{N+1} + c^{\dagger}_{N+1} \right) = - \prod^{N}_{j=1} \left( 1-2 c^{\dagger}_{j} c_{j} \right) \left( c_{N} + c^{\dagger}_{N} \right) \left( c_{1} + c^{\dagger}_{1} \right).
		\end{align}
	\end{subequations}
	One notices that there are two possibilities to hold the above equation. We define $\mathcal{P} \equiv \prod^{N}_{j=1} \left( 1-2 c^{\dagger}_{j} c_{j} \right)$ as a parity operator with eigenvalues $\pm 1$ depending on the total number of fermions occupied (or equivalently the total number of down spins). Since this operator commutes with Hamiltonian $[H ,\mathcal{P}]=0$, we can separate the Hamiltonian into two sectors, as even $H^{(even)}$ and odd $H^{(odd)}$. The first sector (even) has the antiperiodic boundary condition for fermions with the total number being even:
	\begin{subequations}
		\begin{align}
				\label{eqn:antiperiodic}
		\prod^{N}_{j=1} \left( 1-2 c^{\dagger}_{j} c_{j} \right)= 1,\ c_{N+1}= -c_{1}.
		\end{align}
	The other sector has a periodic boundary condition where the total number of fermions is odd:
		\begin{align}
				\label{eqn:periodic}
		\prod^{N}_{j=1} \left( 1-2 c^{\dagger}_{j} c_{j} \right)=- 1,\
		c_{N+1}= c_{1}.
		\end{align}
	\end{subequations}
	With the Jordan-Wigner transformation, we can rewrite the Hamiltonian in terms of the fermion operators as follows,
	\begin{multline}
	\label{eq:HPXY_fermion}
	H_{PXY} =- \sum_{j=1}^{N} \left[ \sum_{l=1}^{N^{(x)}} J_{l}^{(x)} \left( c_{j-1}^{\dagger}c_{j+n_{l}^{(x)}} + c_{j-1}^{\dagger}c_{j+n_{l}^{(x)}}^{\dagger} - 
	c_{j-1}c_{j+n_{l}^{(x)}} - c_{j-1}c_{j+n_{l}^{(x)}}^{\dagger} \right) \right.\\ \left.
	+ \sum_{l'=1}^{N^{(y)}} J_{l'}^{(y)} \left( c_{j-1}^{\dagger}c_{j+n_{l'}^{(y)}} - c_{j-1}^{\dagger}c_{j+n_{l'}^{(y)}}^{\dagger} + 
	c_{j-1}c_{j+n_{l'}^{(y)}} - c_{j-1}c_{j+n_{l'}^{(y)}}^{\dagger} \right) +h(1-2 c^{\dagger}_{j} c_{j}) \right].
	\end{multline}
The above Hamiltonian in terms of fermion operators is also of interest due to recent development in Majorana fermions in the Kitaev's chain~\cite{Kitaev2001}. {But the fermionic Hamiltonian~(\ref{eq:HPXY_fermion}) was obtained from the spin Hamitonian~(\ref{eq:PXY}), and thus, the number of fermions is constrained, related to periodic or antiperiodic boundary condition for fermions.} In Ref.~\cite{Lieb1961}, Lieb, Schutz and Mattis described how to diagonalize such a Hamiltonian. The strategy is to make some transformation (from $c$ fermions to some $\gamma$ fermions) to bring the Hamiltonian in the following diagonal form:
		\begin{equation}
		H=\sum_k \epsilon_k \gamma_k^\dagger \gamma_k + {\rm const.}
		\end{equation}
		Then, the ground state will be obtained by filling up all the modes $k$ that are negative $\epsilon_k<0$, obeying the above fermion number constraints. As we shall see below, it is generally possible to make $\epsilon_k\ge 0$ for most modes, except a few modes that are already diagonal in the $c$-fermion basis (thus, transformation to gamma fermions is not made for them). Moreover, the constraints on the fermion number and the boundary condition on the fermionic operators separate the solutions into two different sectors. Therefore, to determine the ground state, we need to compare the lowest solutions from each sector. Such an issue is important for finite $N$, but can be ignored in the thermodynamic limit.
		An alternative way to solve the Hamiltonian is to go to the Majorana fermion basis, e.g., by letting $\eta_{2j-1}= (c_j + c_j^\dagger)$ and $\eta_{2j}=-i (c_j - c_j^\dagger)$. Then, the Hamiltonian becomes $H=i\sum_{j,j'} A_{j,j'}\eta_j \eta_{j'}$/2, where $A$ is a $2N\times 2N$ real antisymmetric matrix. The matrix $A$ will have spectrum $\{\pm i \epsilon_m\}$ which contains \textit{double} spectrum $\pm\epsilon_m$. However, we will not take the latter approach here.
	
	With the above remarks, let us proceed to diagonalize the Hamiltonian~(\ref{eq:HPXY_fermion}). We use a superscript $(b)$ to indicate which of the two sectors: $b=0$ is for the periodic (odd sector) and $b=1/2$ the antiperiodic (even sector) boundary conditions. As it is translationally invariant, we can perform a Fourier transformation, using
	\begin{subequations}
		\begin{align}
		c_{j} &= \frac{1}{\sqrt{N}} \sum_{k=0}^{N-1} e^{i \frac{2\pi}{N} j(k+b)} \tilde{c}_{k}^{(b)},\\
		\tilde{c}_{k}^{(b)} &= \frac{1}{\sqrt{N}} \sum_{j=0}^{N-1} e^{-i \frac{2\pi}{N} j(k+b)} c_{j}.
		\end{align}
	\end{subequations}
	We then use the identities below (where $x$ and $y$ are integers indexing the sites and the notation $\tilde{}$ \ indicates the operator in the momentum space):
	\begin{subequations}
		\label{eq:gf}
		\begin{align}
		\sum_{j=1}^{N}c_{j+x} c_{j+y} &= \sum_{k=0}^{N-1} e^{i \frac{2\pi}{N} \big[(x-y)(k+b)\big]} \tilde{c}_{k} \tilde{c}_{N-k-2b},\\
		\sum_{j=1}^{N}c_{j+x}^{\dagger} c_{j+y}^{\dagger} &= \sum_{k=0}^{N-1} e^{-i \frac{2\pi}{N} \big[(x-y)(k+b)\big]} \tilde{c}_{k}^{\dagger} \tilde{c}_{N-k-2b}^{\dagger},\\
		\sum_{j=1}^{N}c_{j+x} c_{j+y}^{\dagger} &= \sum_{k=0}^{N-1} e^{i \frac{2\pi}{N} \big[(x-y)(k+b)\big]} \tilde{c}_{k} \tilde{c}_{k}^{\dagger}.
		\end{align}
	\end{subequations}
	Substituting these into Eq. (\ref{eq:HPXY_fermion}), we obtain the following form of the Hamiltonian,
	\begin{align}
	\label{eq:Hamil}
H_{PXY} &=-Nh - \sum_{k=0}^{N-1} \  \left( \sum_l 2 \ J_{l}^{(x)} \cos\Theta^{(x)}_{l}(k) + \sum_{l'} 2 \ J_{l'}^{(y)} \cos\Theta^{(y)}_{l'}(k) - 2h \right) \tilde{c}_{k}^{(b)\dagger} \tilde{c}_{k}^{(b)} \notag \\
	& \quad +  i \left(\sum_l  J_{l}^{(x)} \sin\Theta^{(x)}_{l}(k) -  \sum_{l'} J_{l'}^{(y)} \sin\Theta^{(y)}_{l'}(k) \right) \left[  \tilde{c}_{k}^{(b)} \tilde{c}_{N-k-2b}^{(b)}+ \tilde{c}_{k}^{(b)\dagger} \tilde{c}_{N-k-2b}^{(b)\dagger}
	\right] , \\
	&=-Nh  + \sum_{k=0}^{N-1}  \left[ 2\alpha_k \,\tilde{c}_{k}^{(b)\dagger} \tilde{c}_{k}^{(b)}-i\beta_k\,
	( \tilde{c}_{k}^{(b)} \tilde{c}_{N-k-2b}^{(b)}+ \tilde{c}_{k}^{(b)\dagger} \tilde{c}_{N-k-2b}^{(b)\dagger})\right] \notag,
	\end{align}
We define $\Theta$'s,
	\begin{subequations}
		\begin{align}
		\Theta_{l}^{(x)}(k,b)\equiv\frac{2\pi}{N}(k+b)(1+n_{l}^{(x)}),\\
		\Theta_{l'}^{(y)}(k,b)\equiv\frac{2\pi}{N}(k+b)(1+n_{l'}^{(y)}),
		\end{align}
	\end{subequations}
	and $\alpha$'s, and $\beta$'s to express the solution more compactly,
	\begin{subequations}
		\label{eq:PXY_betalpha}
		\begin{align}
		\beta_{k}^{(b)} &\equiv \sum_{l=1}^{N^{(x)}} J_{l}^{(x)} \sin \Theta^{(x)}_{l}(k,b) - \sum_{l'=1}^{N^{(y)}} J_{l'}^{(y)} \sin \Theta^{(y)}_{l'}(k,b),\\
		\alpha_{k}^{(b)} &\equiv h - \sum_{l=1}^{N^{(x)}} J_{l}^{(x)} \cos \Theta^{(x)}_{l}(k,b) - \sum_{l'=1}^{N^{(y)}} J_{l'}^{(y)}\cos \Theta^{(y)}_{l'}(k,b).
		\end{align}
	\end{subequations}
	We may sometimes suppress the argument $(b)$ in $\Theta$ and the subscript $(b)$ in operators $c$'s, $\alpha$'s and $\beta$'s, when the context is clear.
		We note that in the above, when $\beta_k^{(b)}=0$, the part of the Hamiltonian is already diagonal, i.e., $2\alpha_k^{(b)} \tilde{c}^{(b)\dagger}_k \tilde{c}_k^{(b)}$; we do not need to make any further transformation. This can happen, in the case of $b=0$, with $k=0$ or $k=N/2$ (for $N$ even) and, in the case of $b=1/2$, with $k=(N-1)/2$ for $N$ being odd). We will discuss these below. For $\beta_k^{(b)}\ne 0$, we can diagonalize that part of the Hamiltonian by employing the Bogoliubov transformation that introduces mixing of fermion creation and annihilation operators:
	\begin{subequations}
		\begin{align}
		\tilde{c}_{k} &= \cos \theta_{k} \ \gamma_{k} + i \sin\theta_{k} \gamma_{N-k-2b}^{\dagger},\\
		\tilde{c}_{N-k-2b} &= \cos \theta_{k} \ \gamma_{N-k-2b} - i\sin\theta_{k} \ \gamma_{k}^{\dagger},\\
		\gamma_{k} &= c_{k} \cos \theta_{k}-i \sin \theta_{k} \ c_{N-k-2b}^{\dagger},\\
		\gamma_{N-k-2b} &= c_{N-k-2b} \cos \theta_{k}+i \sin \theta_{k} \ c_{k}^{\dagger},
		\end{align}
	\end{subequations}
	where the Bogoliubov fermions $\gamma$'s obey the same canonical commutation relations: $\{\gamma_{i},\gamma_{j}^{\dagger}\}=\delta_{ij}$.
	By choosing appropriate Bogoliubov angles $\theta_{k}$'s, we can eliminate cross terms $\gamma_k \gamma_{N-k-2b}$ and $\gamma_k^\dagger \gamma_{N-k-2b}^\dagger$, and obtain the diagonalized Hamiltonian:
	\begin{equation}
	H_{PXY} = \sum_{k=0}^{N-1} \epsilon_{k} \left( \gamma_{k}^{\dagger} \gamma_{k}-\frac{1}{2} \right)=\sum_{k\big|\beta_k\ne 0} \epsilon_{k} \left( \gamma_{k}^{\dagger} \gamma_{k}-\frac{1}{2} \right)+\sum_{k\big|\beta_k= 0} 2\alpha_{k} \left( \tilde{c}_{k}^{\dagger} \tilde{c}_{k}-\frac{1}{2} \right),
	\label{eq:ParHam}
	\end{equation}
where $\epsilon_{k}$ (when $\beta_k \neq 0$) is the single Bogoliubov particle's energy spectrum:
	\begin{equation}
	\label{eq:energyspectrum}
	\epsilon_{k}=2 \sqrt{\left( \beta_{k} \right)^{2}+\left( \alpha_{k} \right)^{2}},
	\end{equation}
	and the solution to $\theta_{k}$'s (which we also refer to as the Bogoliubov solution) is given by:
	\begin{subequations}
		\begin{align}
		\tan 2\theta_{k} &= \frac{\beta_{k}}{\alpha_{k}},\\
		\cos2\theta_{k} &= \frac{\left( \alpha_{k} \right)}{\sqrt{\left( \beta_{k} \right)^2 + \left( \alpha_{k} \right)^2}},\\
		\sin\theta_{k} &= \rm{sgn}(\beta_{k}) \ \sqrt{\frac{1-\cos2\theta_{k}}{2}}.
		\end{align}
		\label{eq:ParAng}
	\end{subequations}
When $\beta_k=0$, the part of the Hamiltonian is already diagonal, and thus, it is natural to define $\epsilon_{k\big|\beta_k=0}\equiv 2\alpha_k$, which leads to issue in determining the ground-state configuration in terms of particle occupation. Two key points to consider: (1) there are two sectors $b=0$ (constrained by odd number of fermions) and $b=1/2$ (constrained by even number of fermions), the ground state should have the lowest energy among the two sectors; (2) to obtain the lowest total energy in each sector, we need to consider whether to occupy each $k$ mode or not to make the energy as low as possible. The complication comes when $\epsilon_{k\big|\beta_k=0}\equiv 2\alpha_k$ can become negative in contrast to $\epsilon_{k\big|\beta_k\ne 0}>0$. We elaborate the above points below.
\hfill \break \hfill \break
	\noindent {\bf Subtlety in ground states}. 
	As remarked earlier, we now discuss the subtlety required to obtain the ground state and the energy gap. To attain the true ground state, we have to compare the lowest energy in two sectors: $b=0$ (periodic and odd fermions) and $b=1/2$ (antiperiodic and even fermions). We thus need to make a slight modification to the expression in Eq.~(\ref{eq:energyspectrum}) when $b=0$ and $k=0$ (or equivalently $\Theta(k=0,b=0)=\beta_{0}^{(0)}=0$), since in this case, $k=0$ component in the Hamiltonian~(\ref{eq:Hamil}) is already diagonal:
	\begin{equation}
	\label{eq:k0}
	\epsilon_{k=0}^{(b=0)} \tilde{c}_{0}^{(b=0)\dagger}\tilde{c}_0^{(b=0)}\equiv 2\alpha_{k=0}^{(b=0)}\, \tilde{c}_{0}^{(b=0)\dagger}\tilde{c}_{0}^{(b=0)}.
	\end{equation}
	From the above, it follows that $\gamma_{k=0}^{(b=0)}=\tilde{c}_0^{(b=0)}$ (or equivalently $\theta_{k=0}^{(b=0)}=0$), and thus, Eq.~(\ref{eq:energyspectrum}) for ($k=0$, $b=0$) is modified. Combining constant terms ($Nh$ and others arising from the Jordan-Wigner transformation and commuting $\gamma_k \gamma_k^\dagger=-\gamma_k^\dagger \gamma_k +1$), the contribution from $k=0$ mode becomes $2\alpha_{k=0}\big( \tilde{c}_{0}^{(b=0)\dagger}\tilde{c}_{0}^{(b=0)}-1/2\big)$. Thus, the $\epsilon_{k=0}^{(b=0)}$ reads as $2\alpha_{k=0}^{(b=0)}$ in Eq.~(\ref{eq:energyspectrum}) above. 
	
	Moreover, when $N$ is even, $k$ can take the value $k=N/2$; similarly, the term in the Hamiltonian is also diagonal
	\begin{equation}
	\label{eq:k=N/2}
	\epsilon_{k=N/2}^{(b=0)} \tilde{c}_{N/2}^{(b=0)\dagger}\tilde{c}_{N/2}^{(b=0)}\equiv 2\alpha_{k=N/2}^{(b=0)} \,\tilde{c}_{N/2}^{(b=0)\dagger}\tilde{c}_{N/2}^{(b=0)},
	\end{equation}
	and thus, $\gamma_{k=N/2}^{(b=0)}=\tilde{c}_{k=N/2}^{(b=0)}$ or equivalently $\theta_{k=N/2}^{(b=0)}=0$ (when $N$ is an even integer). The contribution of $k=N/2$ mode to the Hamiltonian becomes $2\alpha_{k=N/2}\big( \tilde{c}_{N/2}^{(b=0)\dagger}\tilde{c}_{N/2}^{(b=0)}-1/2\big)$. Therefore, when $N$ is even, the $\epsilon_{k=N/2}^{(b=0)}$ should be taken as $2\alpha_{k=N/2}^{(b=0)}$ in Eq.~(\ref{eq:energyspectrum}).

Next, we discuss the issues to obtain the lowest total energy. In the $b=0$ sector, the total number of fermions should be odd for the boundary condition in Eq.~(\ref{eqn:periodic}) to be satisfied. {For the number of total sites $N$ being odd, because all excitation $\epsilon_k \ge 0$ (possibly except $\epsilon_{k=0}$), the lowest total energy in this sector has thus exactly one fermion. However, it is {\it not} necessarily that the $k=0$ mode is occupied. This is because when all $\epsilon_k \ge 0$ (including the $k=0$ mode), it is possible that some other mode $k\ne 0$ has the lowest of all, and it is thus energetically favorable to occupy this mode to achieve the lowest total energy, given the constraint of odd number of fermions. } For $N$ being even, the situation can be further complicated by the mode $k=N/2$ with $\epsilon_{k=N/2}=2\alpha_{k=N/2}^{(b=0)}$, which can be negative, and the ground state in this sector may have three fermions. (For such an example, Sect.~\ref{sec:XYhalfway} in the halfway XY model).
	
	According to the above discussions, the associated lowest energy in the $b=0$ sector for even $N$ depends on where it is energetically favorable to occupy, one or three fermions. In the case three fermions are occupied as the lowest energy state, it must involve $\epsilon_{k=0}^{(b=0)}<0$ and $\epsilon^{b=0}_{(k=N/2)}<0$, as well as the lowest of the remaining modes, denoted by $\epsilon_{k'}^{(b=0)}$ (but $\ge 0$). They must satisfy the following condition that 
	\begin{equation}
	\epsilon_{k=0}^{(b=0)}+\epsilon^{(b=0)}_{k=N/2}+\epsilon_{k'}^{(b=0)} < \min\left(\epsilon_{k=0}^{(b=0)},\epsilon^{(b=0)}_{k=N/2}\right).
	\label{eq:threeFermions}
	\end{equation}
	In this case, the lowest energy in this sector is
	\begin{equation}
	E_0^{(b=0, N\, \rm even)}=\epsilon_{k=0}^{(b=0)}+\epsilon^{(b=0)}_{k=N/2}+\epsilon_{k'}^{(b=0)}-\frac{1}{2}\sum_{k=0}^{N-1}\epsilon_k^{(b=0)},
	\label{eq:energyOddThree}
	\end{equation}
	and its associated wave function is
	\begin{equation}
	\ket{\Psi^{(b=0)}}\equiv \tilde{c}^{(0)\dagger}_0 \tilde{c}^{(0)\dagger}_{k=N/2} \tilde{\gamma}^{(b=0)\dagger}_{k'} \prod_{k=1}^{k<\frac{N}{2}}\Big[\cos\theta_{k}^{(0)}
	+i\sin\theta_{k}^{(0)}\,\tilde{c}_k^{(0)\dagger}
	\tilde{c}_{N-k}^{(0)\dagger} \Big]\ket{\Omega}.
	\end{equation}
where $\ket{\Omega}$ denotes the vacuum state. Otherwise,
	\begin{equation}
	\label{eqn:24}
	E_0^{(b=0, N\, \rm even)}=\min_k{\big(\epsilon_{k}^{(b=0)}\big)}-\frac{1}{2}\sum_{k=0}^{N-1}\epsilon_k^{(b=0)}=\epsilon_{k^*}^{(b=0)}-\frac{1}{2}\sum_{k=0}^{N-1}\epsilon_k^{(b=0)},
	\end{equation}
	and the $k^*$ that has the lowest $\epsilon_{k^*}^{(b=0)}$ is often but not necessarily $k=0$ or $k=N/2$; its associated wave function is
	\begin{equation}
	\ket{\Psi^{(b=0)}}\equiv \tilde{\gamma}^{(0)\dagger}_{k^*} \prod_{k=1}^{k<\frac{N}{2}}\Big[\cos\theta_{k}^{(0)}
	+i\sin\theta_{k}^{(0)}\,\tilde{c}_k^{(0)\dagger}
	\tilde{c}_{N-k}^{(0)\dagger} \Big]\ket{\Omega}.
	\end{equation}
	But as $\epsilon_{N-k^*}=\epsilon_{k^*}$, there is a degenerate wave function, by occupying $k=N-k^*$ mode instead.

	When $N$ is odd, the lowest-energy state in this sector necessarily has one fermion, but it does not need to be the $k=0$ mode. The total energy has a similar expression:
	\begin{equation}
	\label{eqn:26}
	E_0^{(b=0, N\, \rm odd)}=\min_k{\big(\epsilon_{k}^{(b=0)}\big)}-\frac{1}{2}\sum_{k=0}^{N-1}\epsilon_k^{(b=0)}.
	\end{equation}
	Similarly, if the minimum $\epsilon_k$ come from $k=0$ mode, then the energy is degenerate.
	
Let us summarize the consideration for the $b=0$ sector. When $N$ is odd, only $\varepsilon_0 = 2 \alpha_0$ may be negative and all other $\varepsilon_{k}\ge 0$, and the odd fermion constraint leads the minimization of total energy to exactly one fermion. On the other hand, when $N$ is even, $k=N/2$ is allowed and $\beta_{k=N/2}=0$. The possibility of $\varepsilon_{k=N/2}=2\alpha_{k= N/2}^{b=0} <0$ and $\varepsilon_0=2\alpha_0<0$ can lead to a 3-fermion configuration having the lowest energy. Hence, we have the possible lowest energies as in Eqs.~(\ref{eq:energyOddThree}), (\ref{eqn:24}) and (\ref{eqn:26}).

	Now we move on to discuss the $b=1/2$ sector. In this sector, the total number of fermions should be even for boundary condition Eq.~(\ref{eqn:antiperiodic}) to be satisfied. When $N$ is odd, the fermion in the mode $k=(N-1)/2$ is not paired with any other mode, and the contribution to the Hamiltonian reads
	$2\alpha_{k=(N-1)/2}\big( \tilde{c}_{N/2}^{(b=1/2)\dagger}\tilde{c}_{(N-1)/2}^{(b=1/2)}-1/2\big)$. That is to say that, when $N$ is odd, $\gamma_{k=(N-1)/2}=c_{k=(N-1)/2}$ (or equivalently $\theta_{k=(N-1)/2}=0$), and thus $\epsilon_{k=(N-1)/2}\equiv 2\alpha_{k=(N-1)/2}$. 
	The lowest energy can arise in two scenarios. First, the simplest case is that there is no fermion. This occurs when 
	\begin{equation}
	\label{eqn:N-1}
	\epsilon_{k=(N-1)/2}+\min_{k\ne (N-1)/2} \epsilon_k \ge 0,
	\end{equation} 
	then 
	\begin{equation}
	\label{eqn:28}
	E_0^{(b=1/2),{\rm (N\, odd)}}= -\frac{1}{2}\sum_{k=0}^{N-1}\epsilon_k^{(b=1/2)}.
	\end{equation}
	But if Eq.~(\ref{eqn:N-1}) is violated with optimal $k'$ (and $N-k'-1$ as well), the ground-state energy in this sector is then degenerate and has the expression
	\begin{equation}
	\label{eqn:29}
	E_0^{(b=1/2),{\rm N\, odd}}=\epsilon_{k=(N-1)/2}+\epsilon_{k'} -\frac{1}{2}\sum_{k=0}^{N-1}\epsilon_k^{(b=1/2)}.
	\end{equation}
	However, there is no such modification when $N$ is even. The lowest energy in the $b=1/2$ sector (with no $\gamma$ fermions occupied) reads:
	\begin{equation}
	\label{eq:energyEven}
	E_0^{(b=1/2),{\rm N\, even}}= -\frac{1}{2}\sum_{k=0}^{N-1}\epsilon_k^{(b=1/2)},
	\end{equation}
	with the associated wavefunction being
	\begin{eqnarray}
	\ket{\Psi^{(b=1/2)}}
	=\prod_{k=0}^{k<\frac{N\!-\!1}{2}}\Big[\cos\theta_{k}
	+i\sin\theta_{k}\,\tilde{c}_k^{\dagger} \tilde{c}_{N-k-1}^{\dagger}
	\Big]\ket{\Omega},
	\end{eqnarray}
	where we suppress the superscript $(b=1/2)$ in $\theta$'s.
	
Let us summarize the discussion for the $b=1/2$ sector. When $N$ is even, all $\beta_k$’s are nonzero and $\varepsilon_k>0$. Therefore, zero fermion has the lowest total energy in that sector. But when $N$ is odd, $\beta_{k=(N-1)/2}=0$ and $\varepsilon_{k=(N-1)/2}=2\alpha_{k=(N-1)/2}$ can be negative, and whether occupying zero or two fermions corresponds to the lowest energy needs a comparison. We have shown the possibilities in Eqs.~(\ref{eqn:28}), (\ref{eqn:29}) and (\ref{eq:energyEven}).
	
	In order to determine the gap above the true ground state, we also need to find the next lowest energy in each sector, in addition to the lowest energies in both sectors $E_0^{(b=1/2)}$ and $E_0^{(b=0)}$. It is {\it not} necessary that the gap is $\Delta=|E_0^{(b=1/2)}-E_0^{(b=0)}|$, even though we find that typically this is the case.
	
	\subsection{Illustrative example: XY model with n-site Z-mediated interaction in the transverse field}
	\label{sec:XnY}
	In this part, we show how to choose parameters and thus obtain the solution of the XY model with $n$-site $Z$-mediated XX and YY interaction. With this model, one can grasp the general features of site-interactions by simply changing $n$ value. For example, the standard XY model can be recovered by taking $n=0$. Let us begin by listing the parameters that characterize this Hamiltonian:
	\begin{subequations}
		\begin{align}
		N^{(x)}&=1,\
		N^{(y)}=1,\\
		J_{l}^{(x)}&=\{(1+r)/2\},\
		J_{l'}^{(y)}=\{(1-r)/2\},\\
		n_{l}^{(x)}&=\{n\},\
		n_{l'}^{(y)}=\{n\}.
		\end{align}
	\end{subequations}
	With the choice of the above parameters, we obtain the corresponding Hamiltonian:
	\begin{equation}
	H_{XnY}= - \sum_{j=1}^{N} \bigg( \frac{1+r}{2} \sigma_{j-1}^{x} \sigma^{z}_{j} \ldots \sigma^{z}_{j+n-1} \sigma_{j+n}^{x} + \frac{1-r}{2} \sigma_{j-1}^{y} \sigma^{z}_{j} \ldots \sigma^{z}_{j+n-1} \sigma_{j+n}^{y} + h \sigma_{j}^{z} \bigg),
	\end{equation}
	which can be diagonalized as 
	\begin{align}
	H &= \sum_{k=0}^{N-1} \epsilon_{k}^{(b)} \left( \gamma_{k}^{(b) \dagger} \gamma_{k}^{(b)}-\frac{1}{2} \right),\\
	\epsilon_{k}^{(b)} &=2 \sqrt{\left( r \sin\phi^{n}_{k} \right)^{2}+\left( h-\cos\phi^{n}_{k} \right)^{2}},
	\end{align}
	with the exceptions of the combination of $b$, $k$ and $N$ mentioned above;
	the solution to the Bogoliubov angles is as follows:
	\begin{equation}
	\tan2\theta_{k} = \frac{r\sin \phi^{n}_{k}}{h-\cos\phi^{n}_{k}},
	\end{equation}
	where we define $\phi_{k}$ for convenience
	\begin{equation}
	\phi^{n}_{k} \equiv \frac{2\pi}{N}(n+1)(k+b),
	\end{equation}
	and $n$ is the number of $\sigma_{z}$ term in each $X$ and $Y$ blocks. The above spectrum $\epsilon_k$, of course, needs to be appropriately modified, for $(k=0,b=0)$, $(k=N/2,b=0)$ for $N$ even and $\big(k=(N-1)/2,b=1/2\big)$ for $N$ odd, etc., as discussed previously. 
	We note that by varying the number of $\sigma_{z}$ one obtains other models:
	\begin{align*}
	n =0 \qquad \rightarrow &\quad \text{XY} \ \text{model,}\\
	n=1\qquad \rightarrow & \quad \text{XY model with three-site interaction ($H_{XzY}$),
	}\\
	n =\frac{N}{2}-1 \qquad \rightarrow & \quad \mbox{(for $N$ even)}\, \textit{halfway interaction.} 
	\end{align*}
	We will investigate quantum phase transitions for these models and others in sections below.
	
	We can also build a different number of Z-mediated sites for each block, such as $(n+2)$-site interaction for $X$ block and $(m+2)$-site interaction for $Y$ block with the following parameters:
	\begin{subequations}
		\begin{align}
		N^{(x)}&=1,\
		N^{(y)}=1,\\
		J_{l}^{(x)}&=\{(1+r)/2\},\
		J_{l'}^{(y)}=\{(1-r)/2\},\\
		n_{l}^{(x)}&=\{n\},\
		n_{l'}^{(y)}=\{m\},
		\end{align}
	\end{subequations}
	and substituting parameters into $H_{PXY}$ gives the following Hamiltonian:
	\begin{equation}
	H_{XnmY}= - \sum_{j=1}^{N} \bigg( \frac{1+r}{2} \sigma_{j-1}^{x} \sigma^{z}_{j} \ldots \sigma^{z}_{j+n-1} \sigma_{j+n}^{x} + \frac{1-r}{2} \sigma_{j-1}^{y} \sigma^{z}_{j} \ldots \sigma^{z}_{j+m-1} \sigma_{j+m}^{y} + h \sigma_{j}^{z} \bigg).
	\end{equation}
	
	\section{\label{sec:level2}Geometric Measure of Entanglement for generalized cluster-XY models}
	
	Entanglement has become a useful tool to study quantum criticality after several pioneering works on the behavior of entanglement near the quantum critical points \cite{Osborne2002,Osterloh2002,Vidal2003,Calabrese2004,Chung2001,Korepin2004}. Many of the previous works on entanglement investigated the domain of bipartite systems. The geometric measurement of entanglement, introduced earlier, was based on a work of Barnum and co-workers \cite{Barnum2001} and developed further by Wei and collaborators~\cite{Wei2003,Wei2005,Orus2008,Orus2008a,Orus2010,Son2011}. 
	
	The main idea of analyzing the geometric entanglement is to find a minimum distance between the entangled state $|\Psi\rangle$ and suitably defined product states, such as 
	\begin{equation}
	\ket{\Phi} \equiv \bigotimes^{n}_{i=1} \ket{\phi^{(i)}}.
	\end{equation}
	An essential quantity is the maximal overlap,
	\begin{equation}
	\Lambda_{\max}(\Psi) \equiv \underset{\Phi}{\max} |\braket{\Phi|\Psi}|,
	\end{equation}
	from which we can define the geometric entanglement
	\begin{equation}
	E_{G}^{(1)}(\Psi) \equiv - \log_{2}\ \Lambda^{2}_{\max}(\Psi),
	\end{equation} 
	and the entanglement density
	\begin{equation}
	\label{eq:geometric_ent}
	\mathcal{E}^{(1)} \equiv \frac{E_{G}^{(1)}(\Psi)}{N},
	\end{equation}
	where $N$ denotes the total number of sites. We note that for GHZ states, $\Lambda_{\max}=1/2$, and thus, $E_G^{(1)}=1$. Similarly by properly defining the product state, we can define the geometric entanglement among blocks with each block containing 2 spins, $E_G^{(2)}$, and its density $\mathcal{E}^{(2)}$, as discussed in Introduction. In the following section, we present derivation of the overlaps for these two scenarios. 
	
	\subsection{Geometric entanglement per site}
	
	Here, we review the derivation of the overlap of the ground state with a product state, comprised of product of single spin states: $\Ket{\Phi_{1}}=(a\Ket{\uparrow}+b\Ket{\downarrow})^{\otimes N}$ which can be written as fermions by applying the Jordan-Wigner transformation
	\begin{subequations}
		\begin{align}
		\Ket{\Phi_{1}}&=\bigotimes_{i=1}^{N}\left( a+b\sigma_{i}^{-}\right) \Ket{\uparrow\uparrow\ldots\uparrow},\\
		&=\prod_{i=1}^{N}\left[ a+b\prod_{j=1}^{i-1}(1-2c_{j}^{\dagger}c_{j})c_{i}^{\dagger}\right] \Ket{\Omega},
		\end{align}
	\end{subequations}
	where $\ket{\Omega}$ is the vacuum with no c fermions. Using this fact, we can further simplify the expression
	\begin{align}
	\Ket{\Phi_{1}}&=\prod_{i=1}^{N} \left[ a+b \ c_{i}^{\dagger}\right] \Ket{\Omega} = a^{N} \prod_{i=1}^{N} e^{b' c_{i}^{\dagger}} \Ket{\Omega},\\
	&=a^N e^{\sum_{i=1}^{N} b' c_{i}^{\dagger}} e^{\sum_{i<j} (b')^2 c_{i}^{\dagger} \ c_{j}^{\dagger}},
	\end{align}
	where we have defined $b' \equiv b/a$. Note that $e^{A}e^{B}=e^{A+B}e^{[A,B]/2}=e^{A+B}e^{A B}$ if $A^2=B^2=0$ and $\{A,B\}=0$. For many such operators, we use $e^{A_{1}}e^{A_{2}}...e^{A_{k}}=e^{\sum A_{i}} e^{\sum_{i<j} A_{i} A_{j}}$ to bring them to the same exponent. Namely, $\prod_{i=1}^{N} e^{b' c_{i}^{\dagger}}= e^{\sum_{i=1}^{N} b' c_{i}^{\dagger}} e^{\sum_{i<j} (b')^2 c_{i}^{\dagger} \ c_{j}^{\dagger}}.$ Next, we need to express $\sum_{i<j} c_{i}^{\dagger} c_{j}^{\dagger}$ in the momentum basis. Notice that we can relax the limit $i < j$ in the sum to $i \leq j$, as $c_{i}^{\dagger} c_{i}^{\dagger}=0$. For simplicity and for the purpose of illustration, we consider quantum XY model with nearest neighbor interaction with $N$ being even and consider the odd sector, i.e., $c_{j+N} = -c_{j}$, and thus, using the following Fourier transformation $c_{j}=\frac{1}{\sqrt{N}}\sum_{k=0}^{N-1}e^{i \frac{2 \pi}{N} j (k+1/2)} c_{k}$, we calculate
	\begin{subequations}
		\begin{align}
		\sum_{j \leq l} c_{j}^{\dagger} c_{l}^{\dagger}&= \frac{1}{N} \sum_{j=1}^{l} \sum_{k,k'=0}^{N-1} e^{-i \frac{2 \pi}{N} j (k+\frac{1}{2}) -i \frac{2 \pi}{N} l (k'+\frac{1}{2})} c_{k}^{\dagger} \ c_{k'}^{\dagger}\\
		&= \frac{1}{N} \sum_{k,k'=0}^{N-1} e^{-i \frac{2 \pi}{N} (k+\frac{1}{2})} \dfrac{e^{-i \frac{2 \pi}{N} l (k'+\frac{1}{2})}-e^{-i \frac{2 \pi}{N} l (k+k'+1)}}{1-e^{-i \frac{2 \pi}{N} (k+\frac{1}{2})}} c_{k}^{\dagger} \ c_{k'}^{\dagger}.
		\end{align}
	\end{subequations}
	Noting that
	\begin{subequations}
		\begin{align}
		\sum_{l=1}^{N} e^{-i \frac{2 \pi}{N} l \ (k'+\frac{1}{2})} &= e^{-i \frac{2 \pi}{N} (k'+\frac{1}{2})} \dfrac{1-e^{-i \frac{2 \pi}{N} N (k'+\frac{1}{2})}}{1-e^{-i \frac{2 \pi}{N} (k'+\frac{1}{2})}}=\dfrac{e^{-i \frac{\pi}{N} (k'+\frac{1}{2})} }{i \sin \left[ \frac{\pi}{N} (k'+\frac{1}{2}) \right]},\\
		\sum_{l=1}^{N} e^{-i \frac{2 \pi}{N} l \ (k+k'+1)} &= N \delta_{k+k'+1,N},
		\end{align}
	\end{subequations}
	we arrive at
	\begin{align}
	\sum_{j \leq l} c_{j}^{\dagger} c_{l}^{\dagger} = \frac{1}{N} \sum_{k,k'=0}^{N-1} \dfrac{e^{-i \frac{\pi}{N} (k+\frac{1}{2})} }{i \sin \left[ \frac{\pi}{N} (k+\frac{1}{2}) \right]} \dfrac{e^{-i \frac{\pi}{N} (k'+\frac{1}{2})} }{i \sin \left[ \frac{\pi}{N} (k'+\frac{1}{2}) \right]} c_{k}^{\dagger} c_{k'}^{\dagger}- \sum_{k=0}^{N-1} \dfrac{e^{-i \frac{\pi}{N} (k+\frac{1}{2})} }{i \sin \left[ \frac{\pi}{N} (k+\frac{1}{2}) \right]} c_{k}^{\dagger} c_{N-k-1}^{\dagger}.
	\end{align}
	The coefficient of the first term is symmetric in $(k,k')$, and thus, the sum makes no contribution, and we can symmetrize the second term, obtaining
	\begin{equation}
	\sum_{j \leq l} c_{j}^{\dagger} c_{l}^{\dagger} = \sum_{k=0}^{N-1} i \cot \left[ \frac{\pi}{N} (k+\frac{1}{2}) \right] c_{k}^{\dagger} c_{N-k-1}^{\dagger}.
	\end{equation}
	Thus, we have rewritten $\ket{\Phi_{1}}$ in terms of fermionic language,
	\begin{equation}
	\ket{\Phi_{1}}= a^N e^{\sum_{i=1}^{N} b' c_{i}^{\dagger}} \ e^{\sum_{i=1}^{N} (b')^2 \sum_{k=0}^{N-1} i \cot \left[ \frac{\pi}{N} (k+\frac{1}{2}) \right] \ c_{k}^{\dagger} c_{N-k-1}^{\dagger}}\Ket{\Omega},	
	\end{equation}
	and we can choose an arbitrary normalizable constant such that $a=\cos \ \frac{\xi}{2}$ and $b=\sin \ \frac{\xi}{2}$. Note that we assume the product state is also translation invariant. 
	
	In many cases in the thermodynamic limit, the ground state is in the sector of $b=1/2$ with no fermion, i.e., $|\Psi_{1/2}\rangle$. So we will illustrate the calculation of the overlap $\langle{\Phi_1}|{\Psi_{1/2}}\rangle$ in order to obtain the entanglement for $|\Psi_{1/2}\rangle$. It is convenient to rewrite $\ket{\Phi_1}$ in the similar pairing form as the ground state for the even $N$ case,
	\begin{equation}
	\ket{\Phi_{1}(\xi)} = \prod_{k=0}^{k<\frac{N-1}{2}} \Big( \cos^{2} \ \frac{\xi}{2}+ i \sin^{2} \ \frac{\xi}{2} \cot \frac{\pi(k+\frac{1}{2})}{N} \ \tilde{c}_{k}^{\dagger}\tilde{c}_{N-k-1}^{\dagger} \Big) \Ket{\Omega},
	\end{equation}
	and thus, we arrive at the overlap for even $N$
	\begin{equation}
	\label{eq:overlap}
	\braket{\Psi_{1/2}|\Phi(\xi)}= \prod_{k=0}^{k<\frac{N-1}{2}}
	\left( \cos\theta_{k}\cos^2\frac{\xi}{2} + \sin\theta_{k} \sin^2\frac{\xi}{2} \cot\frac{\pi(k+\frac{1}{2})}{N} \right).
	\end{equation}
	Maximizing $\log_{2}|\braket{\Psi|\Phi}|^2$ over $\xi$, we obtain geometric entanglement Eq. (\ref{eq:geometric_ent}) and the entanglement density.
	
	One important point of the above calculations is that the product state can be expressed in terms of pair creations from the vacuum, in the same manner as the ground state. We shall see in the next section that for a different type of product states consisting of pairs of sites it is of the form of four-particle creations from the vacuum. Similar to this, the ground state will be conveniently re-expressed as creation of two corresponding pairs to match the structure. In the above calculations of entanglement, we have assumed the ground state has zero Bogoliubov fermions. Similar calculation can be made for ground states having nonzero Bogoliubov fermions, such as that done in Refs. \cite{Wei2005,Wei2011}.	
	
	\subsection{Geometric entanglement per block}
	If we define the product state to be composed of tensor product of states for blocks of spins, we can investigate the geometric entanglement among these blocks as well as the entanglement per block. Each block can consist of $L$ spins. For $L = 2,$ we write product state, where coefficients $a,b,c,d$ below are normalized but arbitrary constants.
	\begin{equation}	
	\ket{\phi^{[2i-1,2i]}}=a\ket{\uparrow}_{2i-1}\otimes \ket{\uparrow}_{2i}+b\ket{\uparrow}_{2i-1}\otimes \ket{\downarrow}_{2i}+c\ket{\downarrow}_{2i-1}\otimes \ket{\uparrow}_{2i}+d\ket{\downarrow}_{2i-1}\otimes \ket{\downarrow}_{2i}.
	\end{equation}
	Using the Jordan-Wigner transformation, we can re-express the total product state $\ket{\Phi}\equiv\otimes_{i=1}^{N/2}\ket{\phi^{[2i-1,2i]}}$ as follows:
	\begin{equation}
	\ket{\Phi}=\bigotimes_{i=1}^{N/2}\left[a+b\prod_{j=1}^{2i-1} (1-2c_{j}^{\dagger} c_{j} ) c_{2i}^{\dagger} + c\prod_{j=1}^{2i-1} (1-2c_{j}^{\dagger} c_{j} ) c_{2i-2}^{\dagger} +d c_{2i-1}^{\dagger} c_{2i}^{\dagger} \right] \ket{\Omega},
	\end{equation}
	where $\ket{\Omega}$ is the vacuum with no $c$ fermions, and we have assumed here that $N$ is even. We note that we have introduced a parameter $c$, which should be clear to distinguish from the operators $c$'s (which carry a site index). Using the fact that the operators $c$'s annihilate the vacuum, we have
	\begin{align}
	\ket{\Phi}&=\bigotimes_{i=1}^{N/2}\left[a+b \ c_{2i}^{\dagger} + c \ c_{2i-2}^{\dagger} +d \ c_{2i-1}^{\dagger} c_{2i}^{\dagger} \right] \ket{\Omega}\\
	&=a^{N/2} \left[\otimes_{i=1}^{N/2} e^{b' \ c_{2i}^{\dagger} + c' \ c_{2i-1}^{\dagger}} \right] e^{d' \sum_{i=1}^{N/2} c_{2i-1}^{\dagger} c_{2i}^{\dagger}}\ket{\Omega},
	\end{align}
	where we have defined $b' \equiv b/a$, $c'\equiv c/a$, and $d'\equiv d/a$. Employing the trick used earlier to bring operators to the same exponent, we arrive at
	\begin{equation}
	\ket{\Phi}=a^{N/2} \ e^{\sum_{i=1}^{N/2} e^{b' \ c_{2i}^{\dagger} + c' \ c_{2i-1}^{\dagger}}} \ e^{\sum_{i<j} (e^{b' \ c_{2i}^{\dagger} + c' \ c_{2i-1}^{\dagger}})(e^{b' \ c_{2j}^{\dagger} + c' \ c_{2j-1}^{\dagger}})} e^{d' \sum_{i=1}^{N/2} c_{2i-1}^{\dagger} c_{2i}^{\dagger}}\ket{\Omega}.
	\end{equation}
	As we also have the two lowest states $\ket{\Psi_{b}} (b=0,1/2)$ expressed in terms of fermionic basis, we can evaluate the overlap $\braket{\Psi_{b}|\Phi}$ in a straightforward, though tedious manner. Note that in the sum $\sum_{i<j}$, we can safely put the limit as $\sum_{i\leq j}$, as when $i = j$, the term vanishes. Thus, we need to evaluate $\sum_{i \leq j}^{N/2} \left(c_{2i}^{\dagger} c_{2j}^{\dagger}, c_{2i-1}^{\dagger} c_{2j-1}^{\dagger}, c_{2i}^{\dagger} c_{2j-1}^{\dagger}, c_{2i-1}^{\dagger} c_{2j}^{\dagger} \right)$, as well as $\sum_{i}^{N/2} c_{2i-1}^{\dagger} c_{2i}^{\dagger}$ in terms of momentum sum. The calculations for $b=1/2$ case are shown as follows:
	\begin{align}
	\sum_{i \leq j}^{N/2} \left( c_{2i-1}^{\dagger} c_{2j}^{\dagger} + c_{2i}^{\dagger} c_{2j-1}^{\dagger}\right) &= -\frac{1}{2} \sum_{k_{1},k_{2}=0}^{N-1} \dfrac{e^{i \frac{2 \pi}{N}(k_{1}+\frac{1}{2})} + e^{i \frac{2 \pi}{N}(k_{2}+\frac{1}{2})}}{1-e^{-i \frac{2 \pi}{N}2(k_{1}+\frac{1}{2})}} e^{-i \frac{2 \pi}{N}2(k_{1}+\frac{1}{2})} \left( \ldots \right) c_{k_{1}}^{\dagger} c_{k_{2}}^{\dagger}\\
	\sum_{i \leq j}^{N/2} c_{2i}^{\dagger} c_{2j}^{\dagger} &= -\frac{1}{2} \sum_{k_{1},k_{2}=0}^{N-1} \dfrac{e^{-i \frac{2 \pi}{N}2(k_{1}+\frac{1}{2})}}{1-e^{-i \frac{2 \pi}{N}2(k_{1}+\frac{1}{2})}} \left( \ldots \right) c_{k_{1}}^{\dagger} c_{k_{2}}^{\dagger}\\
	\sum_{i \leq j}^{N/2} c_{2i-1}^{\dagger} c_{2j-1}^{\dagger} &= -\frac{1}{2} \sum_{k_{1},k_{2}=0}^{N-1} \dfrac{e^{-i \frac{2 \pi}{N}(k_{1}-k_{2})}}{1-e^{-i \frac{2 \pi}{N}2(k_{1}+\frac{1}{2})}} \left( \ldots \right) c_{k_{1}}^{\dagger} c_{k_{2}}^{\dagger}\\
	\sum_{i \leq j}^{N/2} c_{2i-1}^{\dagger} c_{2i}^{\dagger} &= \frac{1}{2} \sum_{k_{1},k_{2}=0}^{N-1} e^{i \frac{2 \pi}{N}(k_{1}+\frac{1}{2})} \left( \ldots \right) c_{k_{1}}^{\dagger} c_{k_{2}}^{\dagger},
	\end{align}
	where $\left( \ldots \right) \equiv \left(\delta_{k_{1}+k_{2}+1,N} + \delta_{k_{1}+k_{2}+1,N/2} + \delta_{k_{1}+k_{2}+1,3N/2}\right)$. There are three Kronecker delta functions, the first of which, $\delta_{k_{1}+k_{2}+1,N}$, represents the same pairing $(k,N-k-1)$ as the ground state. The latter two, $\delta_{k_{1}+k_{2}+1,N/2}$ + $\delta_{k_{1}+k_{2}+1,3N/2}$, however, do not correspond to the same pairing, but instead correspond to terms broken from two pairs of $(k,N-k-1)$ to $(k+ N/2, N/2 -1-k)$ and $(k+ 3N/2, 3N/2 -1-k)$. 
	
	We then collect those quadratic operators in the exponential of $\ket{\Phi}$ in the following form
	\begin{align}
	\hat{O} \equiv \sum_{k=0}^{k<(N/2-1)/2} \left( f_{k} c_{k}^{\dagger} c_{N-k-1}^{\dagger} - f_{N/2-1-k} c_{N/2+k}^{\dagger} c_{N/2-k-1}^{\dagger} + g_{k} c_{k}^{\dagger} c_{N/2-k-1}^{\dagger} + h_{k} c_{k+N/2}^{\dagger} c_{N-k-1}^{\dagger} \right).
	\end{align}
	This division of operators into four groups facilitates the calculation of the overlap. At last, the overlap reads:
\begin{multline}
		\label{eq:block}
		\braket{\Psi_{1/2}|\Phi} =\chi_{N}\prod_{k=0}^{k<(N/2-1)/2} \bigg\{ a^2 \cos\theta_{k} \cos\theta_{\frac{N}{2}-k-1} + d^2 \sin\theta_{k} \sin\theta_{\frac{N}{2}-k-1} \\+ \cos\theta_{\frac{N}{2}-k-1} \sin\theta_{k} \bigg[\frac{b^2+c^2}{2} \cot \frac{2\pi}{N}(k+\frac{1}{2}) +b \ c \cot\frac{2\pi}{N} (k+\frac{1}{2}) \cos\frac{2\pi}{N} (k+\frac{1}{2}) \\+a \ d \sin\frac{2\pi}{N}(k+\frac{1}{2}) \bigg]+ \cos\theta_{k}\sin\theta_{\frac{N}{2}-k-1}\bigg[-\frac{b^2+c^2}{2}\cot\frac{2\pi}{N}(k+\frac{1}{2})\\+b\ c \cot\frac{2\pi}{N}(k+\frac{1}{2})\cos\frac{2\pi}{N}(k+\frac{1}{2})+a \ d\sin\frac{2\pi}{N}(k+\frac{1}{2})\bigg] \bigg\},
		\end{multline}
		with
		\begin{align*}
		\chi_{N}&=1 \ \text{for} ~N/4= \text{integer},\\
		\chi_{N}&=a \cos\theta_{\frac{1}{2}(\frac{N}{2}-1)}+d\sin\theta_{\frac{1}{2}(\frac{N}{2}-1)}~ \text{for}\ N/2=\text{odd integer}.
		\end{align*}
	By maximizing $\log_{2}|\braket{\Psi|\Phi}|^2$ over parameters $a,b,c,d$ we can obtain the entanglement per block. In the thermodynamic limit, it is written as
		\begin{multline}
{\cal E}_2=-\max_{a,b,c,d} 4\int_0^{\pi/2} d\mu \log_2 \bigg\{
	a^2\,\cos\theta(\mu)\cos\theta(\pi-\mu)+ d^2\,\sin\theta(\mu)\sin\theta(\pi-\mu)
		+ \sin[\theta(\mu)-\theta(\pi-\mu)]\frac{b^2+c^2}{2}\cot\mu\bigg.\\ \bigg.+\sin[\theta(\mu)+\theta(\pi-\mu)]
	\big[b\,c\,\cot\mu\cos\mu+a\,d\,\sin\mu\big]\bigg\}.
	\end{multline}
Here we assume the closest \textit{product} state is product of identical two-spin states. 
	
	We note that the above expression will reduce to that for the single-site product states when we set the two-site state 
	\begin{equation}
	a\ket{\uparrow\uparrow}+b\ket{\uparrow\downarrow}+c\ket{\downarrow\uparrow}+d\ket{\downarrow\downarrow}=\left(\alpha\ket{\uparrow}+\beta\ket{\downarrow}\right)\left(\alpha\ket{\uparrow}+\beta\ket{\downarrow}\right),
	\end{equation}
	namely, we set $a=\alpha^2$, $b=c=\alpha\beta$, and $d=\beta^2$.
	In the case of antiferromagnetic ground state, we can no longer assume the single-site product states to be translationally invariant. However, in order to obtain the entanglement per site, we maximize the overlap $\log_{2}|\braket{\Psi|\Phi}|^2$ with the following parameters:
	$a=\alpha\gamma, b=\alpha\delta, c=\beta\gamma, d=\beta\delta$ where $|\alpha|^2+|\beta|^2=|\gamma|^2+|\delta|^2=1$, which comes from a product state of two sites $\underbrace{\left(\alpha\ket{\uparrow}+\beta\ket{\downarrow}\right)}_{\displaystyle \ket{\kappa}}\underbrace{\left(\gamma\ket{\uparrow}+\delta\ket{\downarrow}\right)}_{\displaystyle \ket{\eta}}$.
	
	\section{Examples}
	
	After having given the parameterized exact solutions for the cluster-XY family of Hamiltonians and calculated the overlap for the ground-state entanglement, we now examine a few examples. We did verify numerically that for all the models we consider, the closest product state to the ground state using (1) single-site product ones and (2) two-site product ones can be written as (1) $\ket{\kappa}\ket{\eta}\ket{\kappa}\ket{\eta}...$ and (2) $\ket{\phi^{[1,2]}}\ket{\phi^{[1,2]}}...$, respectively. We also compared numerical exact diagonalization for lowest two energies, indicated by points at the below figures, with our analytic solutions.
	
	\subsection{The anisotropic XY model with three-site interaction (XzY model)}
	\label{sec:XzY}
	The first model analyzed with the geometric entanglement is the celebrated XY model, done in Ref.~\cite{Wei2005}. It was observed that the geometric entanglement displays a singular behavior across the critical line $h_c=1$. This model was also investigated in terms of other entanglement measures, such as the concurrence~\cite{Osborne2002} and the entanglement entropy~\cite{Vidal2003,Calabrese2004,Korepin2004}. The behavior of concurrence is similar. The entanglement entropy shows a logarithmic scaling in the subsystem size at criticality. 
	
	As a first example in our calculation, we present the solution of the anisotropic XY model with three-site interaction (XX and YY, each mediated by one-site Z term) in the transverse field and discuss the ground-state entanglement. Similar Hamiltonians have been examined previously~\cite{Titvinidze2003,Pachos2004,Lou2004,Derzhko2011}, with little emphasis on the entanglement behavior, except for the localizable entanglement in Ref.~\cite{Pachos2004}. This model in one dimension is exactly solvable. We find that near the critical line $h_c=1$, the global entanglement shows divergence and quantum phase transition occurs between a nontrivial SPT phase and a trivial paramagnetic phase. The existence of the continuous transition is also consistent with the behavior of the energy gap. 
	
	The model is characterized by the following parameters, which we introduced earlier, 
	\begin{subequations}
		\begin{align}
		N^{(x)}&=1,\
		N^{(y)}=1,\\
		J_{l}^{(x)}&=\{(1+r)/2\},\
		J_{l'}^{(y)}=\{(1-r)/2\},\\
		n_{l}^{(x)}&=\{1\},\
		n_{l'}^{(y)}=\{1\}.
		\end{align}
	\end{subequations}
	
			\begin{figure}[t]
		\centering
		\includegraphics[width=1\textwidth]{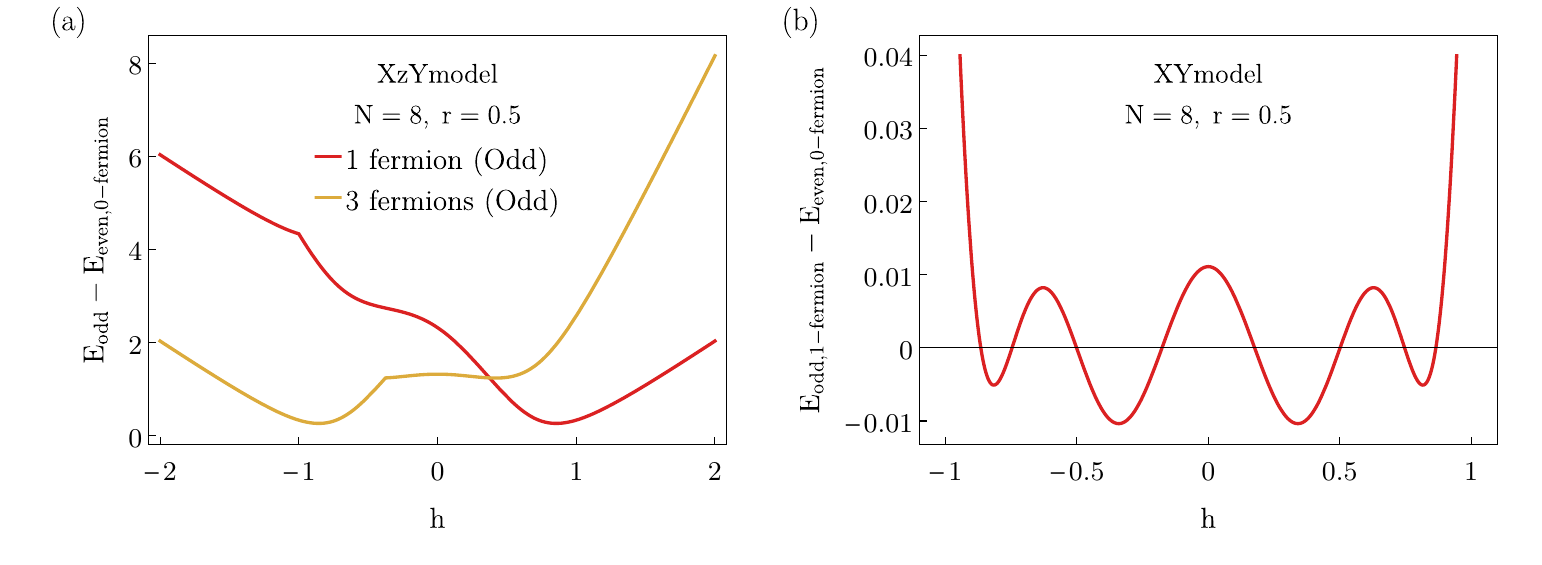}
		\caption{\label{fig:dEoddEeven}
			The energy difference: $E_{\rm odd}-E_{\rm even}$, where $E_{\rm odd}$ is the lowest energy in the odd sector and $E_{\rm even}$ is that in the even sector. (a) For the XzY model with $r=0.5$ and $N=8$. It is seen that the ground-state energy is always $E_{\rm even}$, from the even sector. (b) For XY model with $r=0.5$ and $N=8$. In contrast, it is seen that the ground state switches back and forth between the even and odd sectors, depending on the value of $h$. 
		}
	\end{figure}
	Substituting these terms into $H_{PXY}$ (\ref{eq:PXY}), we obtain the XzY model in the transverse field:
	\begin{align}
	\label{eqn:HXzY}
	H_{XzY}&= - \sum_{j=1}^{N} \left[ \frac{1+r}{2} \sigma_{j-1}^{x} \sigma_{j}^{z} \sigma_{j+1}^{x} + \frac{1-r}{2} \sigma_{j-1}^{y} \sigma_{j}^{z} \sigma_{j+1}^{y} + h \sigma_{j}^{z} \right],
	\end{align}
	where $r$ is a magnetic anisotropy constant between $\sigma_{x}$ and $\sigma_{y}$ terms with $0 \le r \leq 1$. When $r=1$ (the Ising limit), the model reduces a cluster model \cite{Doherty2009}, and in the limit $r=0$, it becomes an isotropic XY model with three-site interaction.
	Using Eq. (\ref{eq:PXY_betalpha}), we calculate $\alpha_{k}$ and $\beta_{k}$:
	\begin{align}
	\beta_{k} &= \Big(\frac{1+r}{2}\Big) \sin \Theta^{(x)}_{l} - \Big(\frac{1-r}{2}\Big) \sin \Theta^{(y)}_{l'},\\
	\alpha_{k} &= {h - \Big(\frac{1+r}{2}\Big) \cos\Theta^{(x)}_{l} -\Big(\frac{1-r}{2}\Big) \cos \Theta^{(y)}_{l'}},
	\end{align}
	with $
	\Theta_{1}=\Theta^{(x)}_{1}=\Theta^{(y)}_{1}=\frac{4\pi}{N}(k+b)$. We then obtain the diagonalized Hamiltonian and the exact energy spectrum (see Eqs. \ref{eq:ParHam}-\ref{eq:k=N/2}):
	\begin{align}
	H &= \sum_{k=0}^{N-1} \epsilon_{k}^{(b)} \left( \gamma_{k}^{(b) \dagger} \gamma_{k}^{(b)}-\frac{1}{2} \right).
	\end{align}
	\begin{figure*}[t]
		\includegraphics[width=1\textwidth]{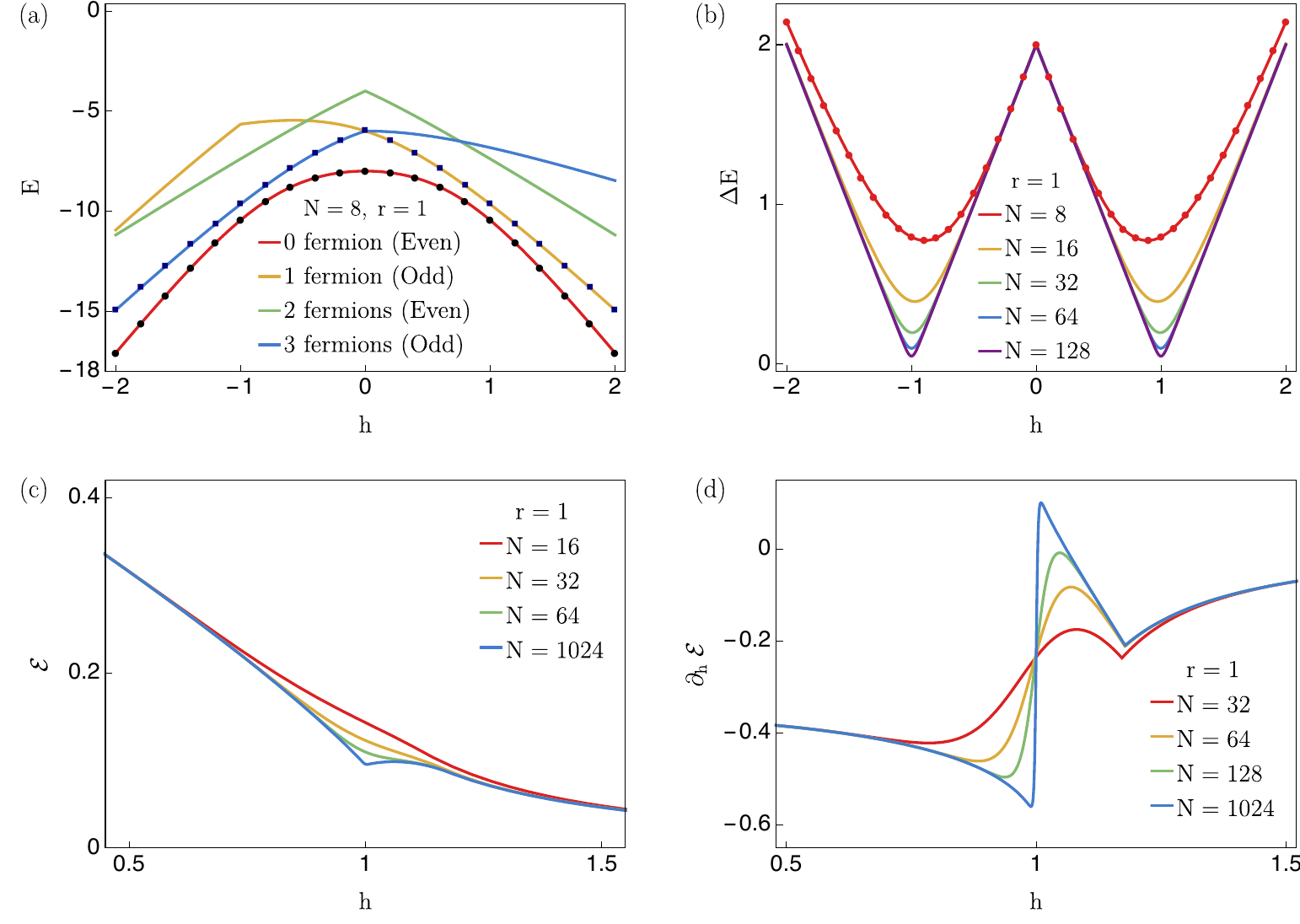}
		\caption{(a) {Lowest few energy levels vs. $h$ for the XzY model with an anisotropy $r=1$ and the system size $N=8$. The model essentially becomes the Ising model with next-nearest neighbor interaction (except the mediating $Z$ factor) in the transverse field. In the odd sector, lowest one-fermion and three-fermions energy levels intercept at $h=0$. The red line (bottom line) indicates the ground state comes from the even sector with zero fermion occupation. (b) The energy difference between the ground and first excited states as a function of $h$ at $r=1$. At the critical point $h=1$, the energy gap is closing as a function of the system size, which indicates a second-order quantum phase transition. In the thermodynamic limit ($N\rightarrow\infty$), the energy gap becomes $2\big| 1-|h| \big|.$ We show that numerical exact diagonalization for lowest two energies (points) and our analytic solutions (curves) agree. (c) Quantum entanglement of the XzY model with the anisotropy $r=1$ and with increasing system sizes $N= 16, 32, 64, 1024$ (from top to bottom). (d) The derivative of the entanglement density of the XzY model for $r=1$, where $N= 32, 64, 128, 1024$ (from top to bottom for $h<1$). The derivative of entanglement diverges and the QPT occurs at $h=1$ between a nontrivial SPT phase for $h<1$ and a trivial paramagnetic phase for $h>1$}. }
		\label{fig:XzYr1}
	\end{figure*}
	The eigenvalues can be obtained by carefully analyzing odd ($b=0$, periodic boundary conditions) and even ($b=1/2$, antiperiodic boundary conditions) sectors separately, assuming $N$ is even or odd, respectively:
	\begin{align}
	\epsilon_{k}^{(b)}= \left\{\begin{array}{lr}
	2 (h-1), & \text{for } k=0 \wedge b=0\\
	2 (h-1), & \text{for } k=\frac{N}{2} \wedge b=0\\
	2 (h-1), & \text{for } k=\frac{N-1}{2} \wedge b=1/2
	\end{array}\right\} = 2 \alpha_{k}^{(b)},
	\end{align}
	or otherwise ($N$ can be either even or odd):
	\begin{align}
	\epsilon_{k}^{(b)}&= 2 \sqrt{\left( \beta_{k} \right)^{2}+\left( \alpha_{k} \right)^{2}}= 2 \sqrt{\left( r \sin \frac{4\pi}{N}(k+b)\right)^{2}+\left( h-\cos \frac{4\pi}{N}(k+b)\right)^{2}},
	\end{align}
	with the corresponding Bogoliubov solution:
	\begin{align}
	\tan 2\theta_{k}^{(b)} &= \frac{\beta_{k}}{\alpha_{k}} = \frac{r \sin \Theta_{1} }{h - \cos \Theta_1}.
	\end{align}

	\begin{figure*}[t]
		\includegraphics[width=1\textwidth]{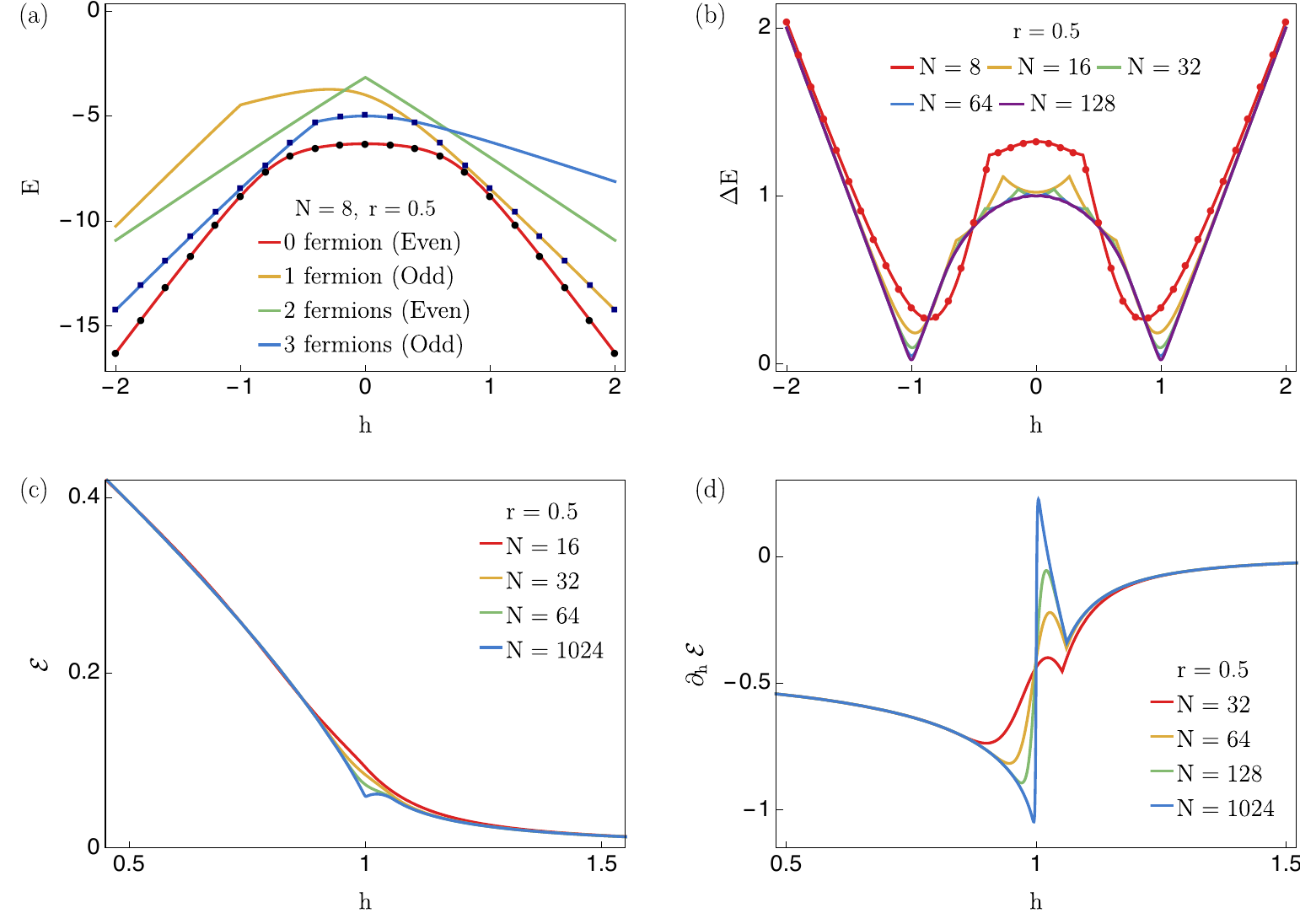}
		\caption{(a) {Lowest few energy levels vs. $h$ for the XzY model with an anisotropy $r=0.5$ and the system size $N=8$. In the specific region ($h\lesssim 0.5$), the first excited state has three-fermion occupation, which is energetically favorable then one-fermion occupation. The possibility of this peculiarity has been discussed in Eq. (\ref{eq:threeFermions}). However, the red line (bottom line) shows that the ground state energy comes from the even sector with zero fermion occupation. We compare numerical exact diagonalization for lowest two energies (points) with our analytic solutions (curves). (b) The energy difference of the ground and first excited states as a function of $h$ at $r=0.5$. At the critical point $h=1$, the energy gap is closing as a function of the system size, which indicates a second order quantum phase transition. (c) Quantum entanglement of the XY model with three-site interaction in the transverse field (also labeled as the XzY model), where the anisotropy $r=0.5$ with increasing system sizes $N= 16, 32, 64, 1024$ (from top to bottom). (d) The derivative of entanglement density of the XzY model for $r=0.5$ and $N= 32, 64, 128, 1024$ (from top to bottom for $h<1$). The derivative of entanglement diverges and the QPT occurs at $h=1$ between a nontrivial SPT phase for $h<1$ and a trivial paramagnetic phase for $h>1$.}}
		\label{fig:XzYr05}
	\end{figure*}

	One notices that the solution is similar to the solution of the standard XY model~\cite{Lieb1961,DePasquale2009,Wei2011}. The only difference occurs in the momentum space by a factor of two, i.e., in the XY model $\Theta_1$ is $2\pi(k+b)/N$ instead of $4\pi(k+b)/N$. But there are some differences that are related to the subtlety in getting the global lowest energy state. For instance, in the XY model with $r\ne 1$, the state of the lowest energy can come from either the even or the odd sector, as illustrated in Fig.~\ref{fig:dEoddEeven}(b) for $r=0.5$. As a function or $h$, the ground state switches between the two sectors, as the lowest energy changes between $E_0^{(b=0)}$ and $E_0^{(b=1/2)}$. But for the XzY model, the ground state is always in the even sector with zero fermion, as illustrated in Fig.~\ref{fig:dEoddEeven}(a). Moreover, for the odd-number fermion case ($b=0$), the lowest-energy level in this sector depends on the field parameter ($h$) and the anisotropy constant ($r$). For example, in the Ising limit where $r=1$, the odd sector has three-fermion occupation as the lowest-energy state in the region of $h<0$; otherwise, it is energetically favorable to occupy one fermion for even $N$; see Fig.~\ref{fig:XzYr1}(a) and also Fig.~\ref{fig:dEoddEeven}(a). However, the true ground state arises from the $b=1/2$ (even) sector and has no $\gamma$ fermion. This phenomenon differs from the standard XY model, where the lowest energy in the odd sector always has one-fermion occupation. The possibility of such peculiarity is discussed in the Sect.~\ref{sec:par}; see discussions around Eq.~(\ref{eq:threeFermions}). We note that for a finite system size $N$ (even) and $r=0.5$, the lowest-energy level in the odd sector has three fermions from negative $h$ values up to about $h\approx0.4$; see Figs.~\ref{fig:XzYr05}(a) and ~\ref{fig:dEoddEeven}(a). Moreover, the energy gap between the ground and the first excited state is closing with an increasing system size $N$ at $h=1$, implying a quantum phase transition there; see Figs. \ref{fig:XzYr05}(b) and \ref{fig:XzYr1}(b). For small finite sizes, the gap as a function of $h$ is not smooth for $r=0.5$. In contrast, the gap vs. $h$ is smooth for $r=1$ even with finite sizes, and {in the thermodynamic limit $N\rightarrow\infty$, the energy gap for $r=1$ (Ising limit of the XzY model) becomes $2\big| 1-|h| \big|$.}
	
		\begin{figure}[t]
		\centering
		\includegraphics[width=.7\textwidth]{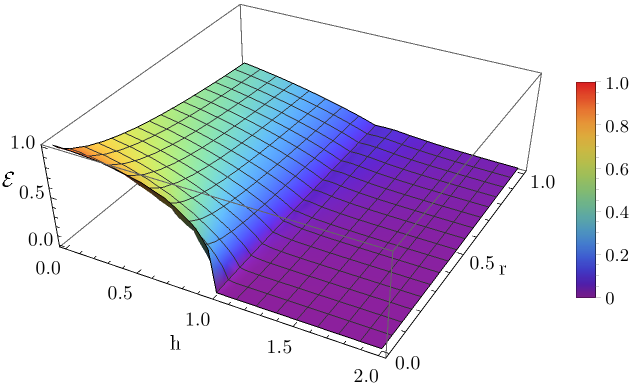}
		\caption{\label{fig:3dEntXzY} Entanglement density per site vs transverse magnetic field ($h$) vs anisotropy ($r$) for XzY model with $N=1000$ spins.}
	\end{figure}		
	
	To examine the quantum phase transition in the phase diagram, we also calculate geometric entanglement and we plot the entanglement per site in Fig.~\ref{fig:3dEntXzY} over a wide range of $r$ and $h$. It is visible that the behavior of entanglement is singular across $h=1$, similar to that in the standard XY model~\cite{Wei2005}. We illustrate this for two different $r$'s ($r=0.5$ and $r=1$) in Figs. \ref{fig:XzYr05}(c) and \ref{fig:XzYr1}(c), as well as the entanglement derivative w.r.t. $h$ in Figs. \ref{fig:XzYr05}(d) and \ref{fig:XzYr1}(d). {The derivative of the entanglement develops singularity, which indicates a quantum phase transition.}
	
	From the above, it follows that for $r=1$ the Hamiltonian reduces to 
	\begin{equation}
	H=-\sum_j (\sigma_{j-1}^x \sigma_{j}^z\sigma_{j+1}^x + h \sigma_j^z).
	\end{equation}
	The model has a $Z_2\times Z_2$ symmetry, generated by $U_e=\prod_{j\, \rm even} \sigma_j^z
	$ and $U_o=\prod_{j\, \rm odd} \sigma_j^z$~\cite{Montes2012}. At $h=0$, the ground state is known to be the cluster state, which is a nontrivial SPT state. (One expects this nontrivial SPT order to hold for general $n$-site mediated Ising model with $Z_2^{\otimes n+1}$ symmetry; see Ref.~\cite{Lahtinen2015}.) At large $h$, the ground state is a trivial paramagnetic state. As we have seen that there is a quantum phase transition at $h=1$, detected by the gap closing and the entanglement singularity, the SPT order appears in the region $|h|\le 1$. In fact, XzY model Eq.~(\ref{eqn:HXzY}) at any $r$ has the $Z_2\times Z_2$ symmetry, and we expect that for $0<r\le 1$, the phase diagram contains a nontrivial SPT phase for $h<1$ (as there is no phase transition inside that region) and a trivial paramagnetic phase for $h>1$, separated at a critical line at $h=1$. The reason $r=0$ line is excluded is because the system is gapless for $h\in[0,1]$ at $r=0$. This compares to the standard XY model, where $h=1$ separates a ferromagnetic phase from a paramagnetic phase. From the results in Ref.~\cite{Lahtinen2015}, we also expect that this is generic behavior for general but finite $n$ (where the interaction is restricted to be short-ranged).
	
	\subsection{XY model with halfway interaction}
	\label{sec:XYhalfway}
		
	\begin{figure}[t]
		\centering 
		\includegraphics[width=1\textwidth]{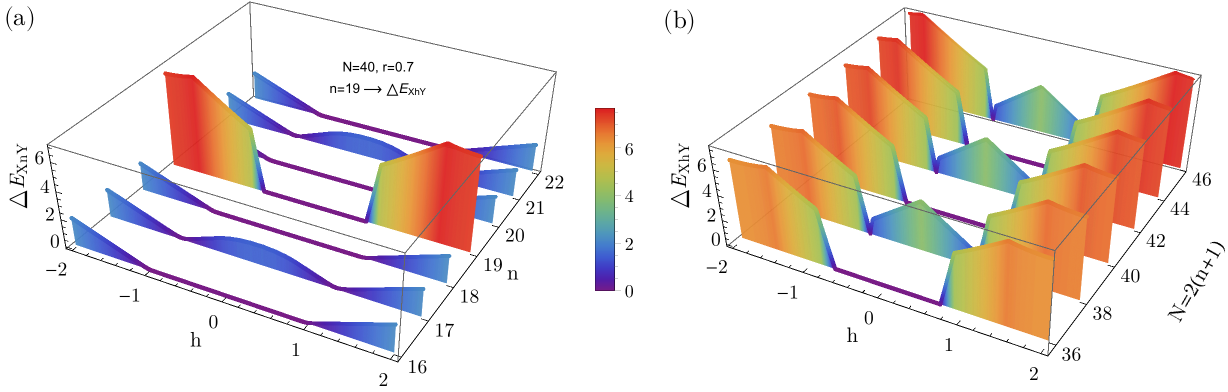}
		\caption{\label{fig:halfway}
			{ (a) The figure illustrates the energy gap for XnY model (namely, XY model with $n$-site Z mediated interaction) with the anisotropy $r=0.7$ at a fixed system size $N=40$ vs. the mediating Z number ($n$) between 16-22 and vs. the transverse magnetic field $h$. We notice a jump in the energy gap at $\sqrt{1-r^2}\approx0.714$ for the halfway XY model ($n=N/2-1=19$). (b) The right figure illustrates the energy gap for the halfway XY model (denoted by XhY) with the following parameters: $N^{(x)}=N^{(y)}=1,J_{l}^{(x)}=\{(1+r)/2\},J_{l'}^{(y)}=\{(1-r)/2\}, n_{l}^{(x)}=n_{l'}^{(y)}=\{N/2-1\}.$ The energy gap has different characteristics between $N=4m$ and $N=2(2m+1)$, as the former is degenerate and the latter is gapped in the region $|h|<\sqrt{1-r^2}.$ 
			}
		}
		
	\end{figure}
	
	\begin{figure}[t]
		\includegraphics[width=1\textwidth]{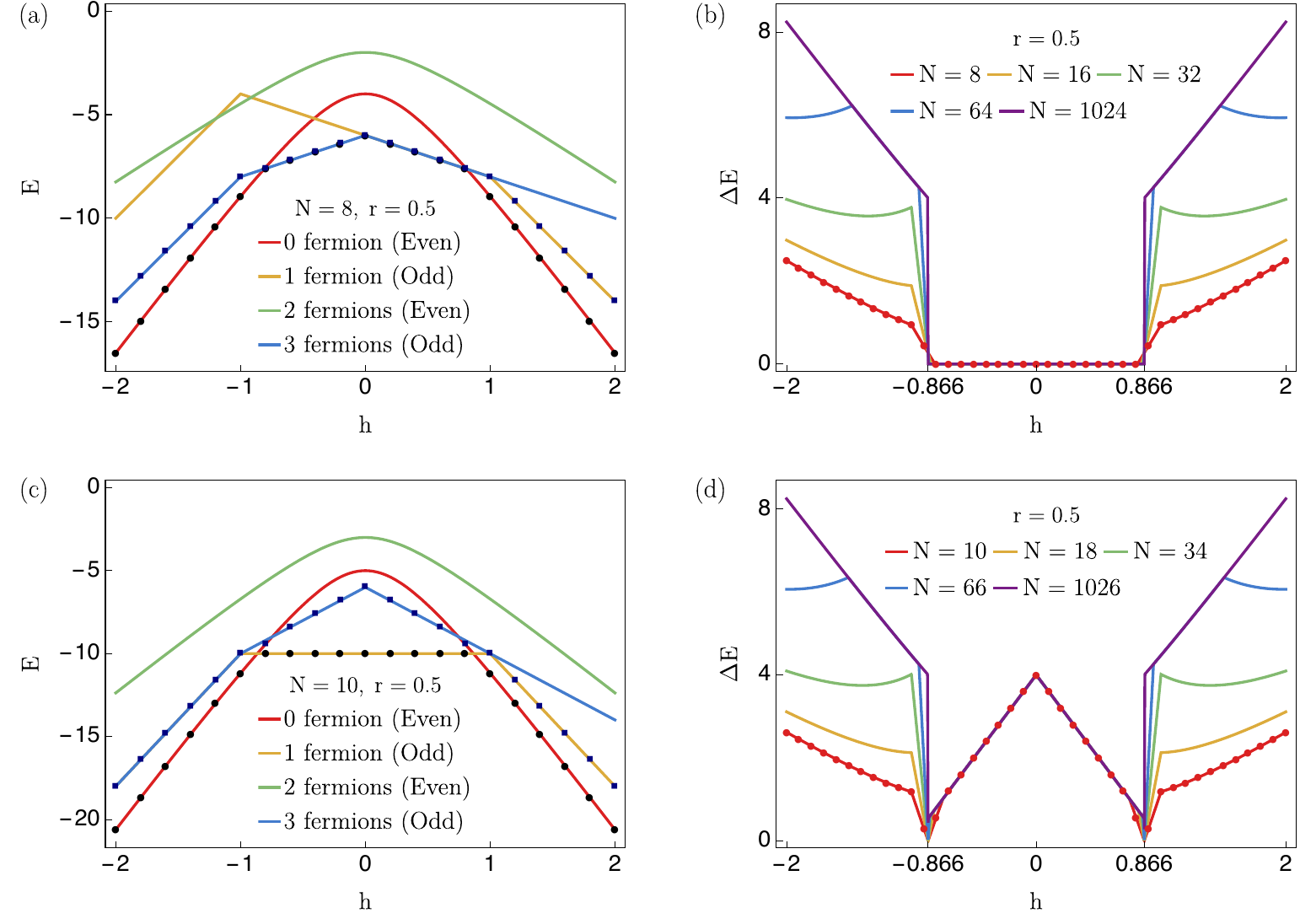}
		\caption{{Lowest few energy levels vs. $h$ for the halfway XY model at $r=0.5$ with top (a): $N=8$; bottom (c): $N=10$. This model shows that the ground state change from the odd to even sector at the transition. The right panel illustrates the energy gap vs. $h$ for the halfway XY model at $r=0.5$. Top (b): $N=4m$; bottom (d): $N=2(2m+1)$. There is clearly a difference between $N=4m$ and $N=2(2m+1)$. In the former, it is gapless in the range $-0.86\lesssim h\lesssim 0.86$, but has a jump to a finite gap outside that range. On the other hand, in the latter case of $N=2(2m+1)$, inside the region $-0.86\lesssim h\lesssim 0.86$, it is gapped, but the size of the gap has a jump at $h\approx \pm 0.86$. This suggests that the transition there is first-order, consistent with the level crossing, shown in (a) and (c)}. We confirm our analytic solutions (curves) with the results obtained from numerical exact diagonalization for lowest two energies and energy gap (points).
		}
		\label{fig:GapHalf_r05}
	\end{figure}
	
	In Sect. \ref{sec:XnY}, we introduced an illustrative example of the XY model with $n$-site $Z$-mediated XX and YY interactions. For $n=0$ and $n=1$, we recover the standard XY model and the XY model with three-site interaction (XzY model) investigated in the previous example. In this part, we demonstrate how a specific choice of site interaction, $n=N/2-1$ (\textit{halfway interaction}) exhibits different behavior from that of $n=0,1$, and has no quantum phase transition at $h=1$. This is a rather interesting result since except at this arbitrary point ($n\neq N/2-1$), the XY model generically exhibits a quantum phase transition for each $n$-site interaction, as seen by vanishing of the gap there in Fig.~\ref{fig:halfway}(a). {Moreover, we also discover a first-order phase transition in the XY model with halfway interaction in the region of $0 \leq r<1$. (The halfway interaction only occurs for even system sizes $N$). In this limit, the first-order transition occurs at the Barouch-McCoy circle \cite{barouch1971}, namely $r^2+h^2=1$. For example, in the case of $r=0.7$ the phase transition occurs at $h_{c}=\sqrt{1 - 0.7^2}\approx 0.714$ as illustrated in Fig.~\ref{fig:halfway}(b).} 
	We note that there is an even-odd effect in $N/2$ and the behavior of the gap is different. 

	We note that for the standard XY model, the Barouch-McCoy circle represents only a crossover that divides the ferromagnetic phase into two regions. Here, for the halfway interaction, the circle represents a curve of first-order transition points.
	
	First, let us define the parameters that give the XY model with $n$-site interaction
	\begin{subequations}
		\begin{align}
		N^{(x)}&=1,\
		N^{(y)}=1,\\
		J_{l}^{(x)}&=\{(1+r)/2\},\
		J_{l'}^{(y)}=\{(1-r)/2\},\\
		n_{l}^{(x)}&=\{n\},\
		n_{l'}^{(y)}=\{n\},
		\end{align}
	\end{subequations}
	yielding the corresponding Hamiltonian:
	\begin{equation}
	H_{XnY}= - \sum_{j=1}^{N} \bigg( \frac{1+r}{2} \sigma_{j-1}^{x} \sigma^{z}_{j} \ldots \sigma^{z}_{j+n-1} \sigma_{j+n}^{x} + \frac{1-r}{2} \sigma_{j-1}^{y} \sigma^{z}_{j} \ldots \sigma^{z}_{j+n-1} \sigma_{j+n}^{y} + h \sigma_{j}^{z} \bigg).
	\end{equation}
	This Hamiltonian can be diagonalized into the form Eq.~(\ref{eq:ParHam}) and we obtain the following Bogoliubov solution (with $\phi^{n}_{k} \equiv \frac{2\pi}{N}(n+1)(k+b)$):
	\begin{equation}
	\tan2\theta^{(b)}_{k} = \frac{r\sin \phi^{n}_{k}}{h-\cos\phi^{n}_{k}}.
	\end{equation}
	In the case of halfway interaction, we substitute $n=N/2-1$ to simplify Bogoliubov solution, respectively, for the even ($b=1/2$) and the odd sector ($b=0$):
	\begin{subequations}
		\begin{align}
		\tan2\theta^{(1/2)}_{k} &= \frac{r\sin \Big[\pi \left( k+\frac{1}{2}\right) \Big] }{h-\cos\Big[\pi \left( k+\frac{1}{2}\right) \Big]}= \frac{(-1)^{k} r}{h},\\
		\tan2\theta^{(0)}_{k} &= \frac{r\sin \left( \pi k\right) }{h-\cos\left(\pi k\right)}=0,
		\end{align}
	\end{subequations}
	with following energy spectrum for {odd} $N$ and $b=1/2$, and for {even} $N$ and $b=0$:
	\begin{align}
	\epsilon_{k}^{(b)}= \left\{\begin{array}{lr}
	2 (h-1), & \text{for } k=0 \wedge b=0\\
	2 \left[h-(-1)^{N/2}\right], & \text{for } k=\frac{N}{2} \wedge b=0\\
	2 \ h, & \text{for } k=\frac{N-1}{2} \wedge b=1/2
	\end{array}\right\} = 2 \alpha_{k}^{(b)},
	\end{align}
	or otherwise:
	\begin{align}
	\epsilon_{k}^{(b)}&= 2 \sqrt{\Big[h-\cos \big(\pi (b+k)\big)\Big]^2+\Big[r \sin \big(\pi (b+k)\big)\Big]^2}= \left\{\begin{array}{lr}
	2 |h-(-1)^k|, & \text{for } b=0,\\
	2 \sqrt{h^2+r^2}, & \text{for } b=1/2.
	\end{array}\right.
	\end{align}
	
	\begin{figure*}[t]
		\includegraphics[width=1\textwidth]{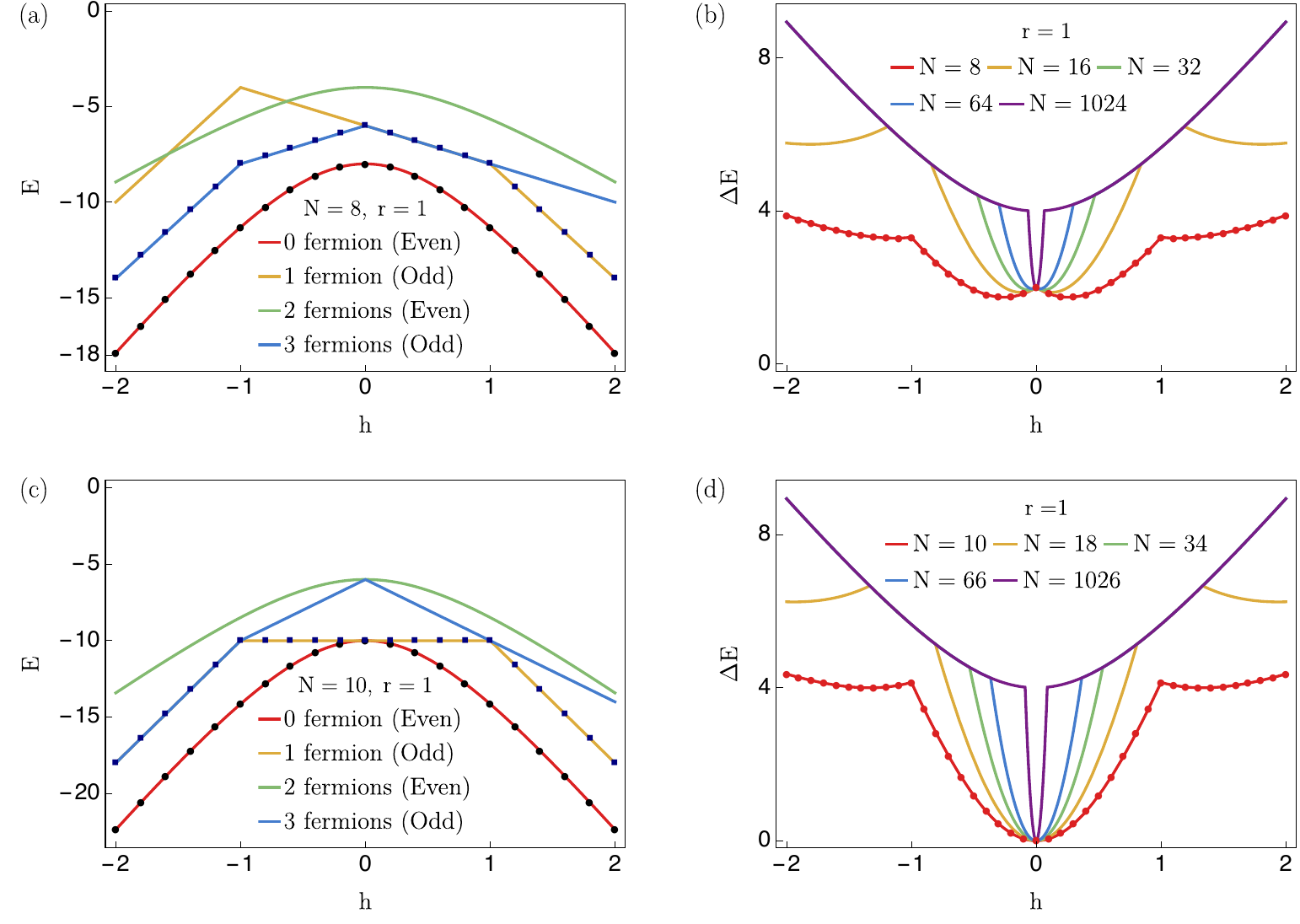}
		\caption{The lowest two levels for even and odd sectors with top (a): $N=8$; bottom (c): $N=10$, for the halfway Ising model at $r=1$. We note that for the negative $h$, three-fermion occupation occurs as the lowest level in the odd sector, instead of one-fermion, which satisfies the inequality is shown in Eq. (\ref{eq:threeFermions}). The true ground state is constructed by the even sector ($b=1/2$) with no fermion. The right panels illustrate the energy gap between the ground state and the first excited state for halfway XY model as a function of $h$ at $r=1$. Top (b): $N=4m$; bottom (d): $N=2(2m+1)$. We see that as $N$ becomes very large, the system becomes gapped at all $h$, except possible double degeneracy at $h=0$. This shows that there is no phase transition in the thermodynamic limit. We show that numerical exact diagonalization for lowest two energies (points) and our analytic solutions (curves) agree.
		}
		\label{fig:GapHalf_r1} 
	\end{figure*}
	To obtain the ground state and the first excited state, one should examine even and odd sectors carefully. This model shows vacua competition \cite{DePasquale2009} similar to the standard XY model, meaning that odd and even sectors switch the roles of being the true ground state depending on $h$. This competition is lifted in the Ising limit where $r=1$ and the ground state is certainly constructed from the even sector ($b=1/2$) with no fermion; except when $N=2(2m+1)$ and at $h=0$, another degenerate ground state is from the odd sector with one fermion; see Fig.~\ref{fig:GapHalf_r1}. In the case of $r=0.5$, the switching happens around $h\approx 0.866$. The ground state becomes dominated by the odd sector in the range $-0.87 \lesssim h \lesssim 0.87$, but outside that range the ground state comes from the even sector ($b=0$) with zero-fermion occupation; see Fig.~\ref{fig:GapHalf_r05}. In particular, for $-0.87 \lesssim h < 0$ and with $N=4m$, the lowest-energy level in the odd sector has three-fermion occupation instead of one fermion, as it is energetically favorable to occupy three fermions in the {odd sector} rather than just one fermion. In fact, in this region, the ground state is degenerate \big(not shown explicitly in Fig.~\ref{fig:GapHalf_r05}(a), but is shown in Fig.~\ref{fig:GapHalf_r05}(b)\big), both degenerate ground states have 3 fermions. But in {$0\leq h \lesssim 0.87$}, the lowest one-fermion and three-fermion states become degenerate. For $N=2(2m+1)$ and $-0.87 \lesssim h \lesssim 0.87$, the lowest energy is dominated by the one-fermion state in the odd sector. This phenomenon is anticipated earlier in Eqs. (\ref{eq:k0}-\ref{eq:energyOddThree}). Using these equations, we also calculate the lowest energy for the odd/even sector and the true energy gap which is shown in the Fig.~\ref{fig:GapHalf_r05}. All of these suggest that there is a first-order phase transition for the halfway XY model with $0\le r<1$, as the transition is due to a level crossing. However, for $r=1$, the halfway Ising model, the gap closes at $h=0$ only for $N=2(2m+1)$, but not for $N=4m$.
	
	There is an interesting picture that emerges. In the standard XY model in a transverse field, there is a crossover curve, the so-called Barouch-McCoy circle, given by $r^2+h^2=1$ \cite{barouch1971}. The crossover curve divides the ferromagnetic phase into two regions: (i) inside the arc, the spin-spin correlation functions display oscillatory behavior, and (ii) outside the arc, the correlation functions have no oscillatory behavior. On the arc, the ground state is essentially a product state, also detected by zero geometric entanglement previously in Ref.~\cite{Wei2005}. Here for the halfway XY model, the crossover arc, $r^2+h^2=1$ is promoted to a first-order transition curve, due to the mediated long-range Z string of a specific length $n=N/2-1$. Thus, the transition field $h$ for $r=0.5$ is $h_c(r=0.5)=\sqrt{1-0.5^2}\approx 0.886$, agreeing with our calculations of the energy gap in Fig.~\ref{fig:GapHalf_r05}. This works for other value of $0\le r<1$ as well, {see Fig. \ref{fig:halfway}(b) for $r=0.7$ case}. The behavior of the $r=1$ halfway Ising model is different, as there is a closing of the energy gap at $h=0$ only for the total site number being $N=2(2m+1)$, as shown in Fig.~\ref{fig:GapHalf_r1}. But in the thermodynamic limit, the energy gap $\Delta E$ is always finite, except at the peculiar point $h=0$, namely that it does not close continuously. We thus do not regard this as a phase transition.
	
\begin{figure*}[t]
		\includegraphics[width=1\textwidth]{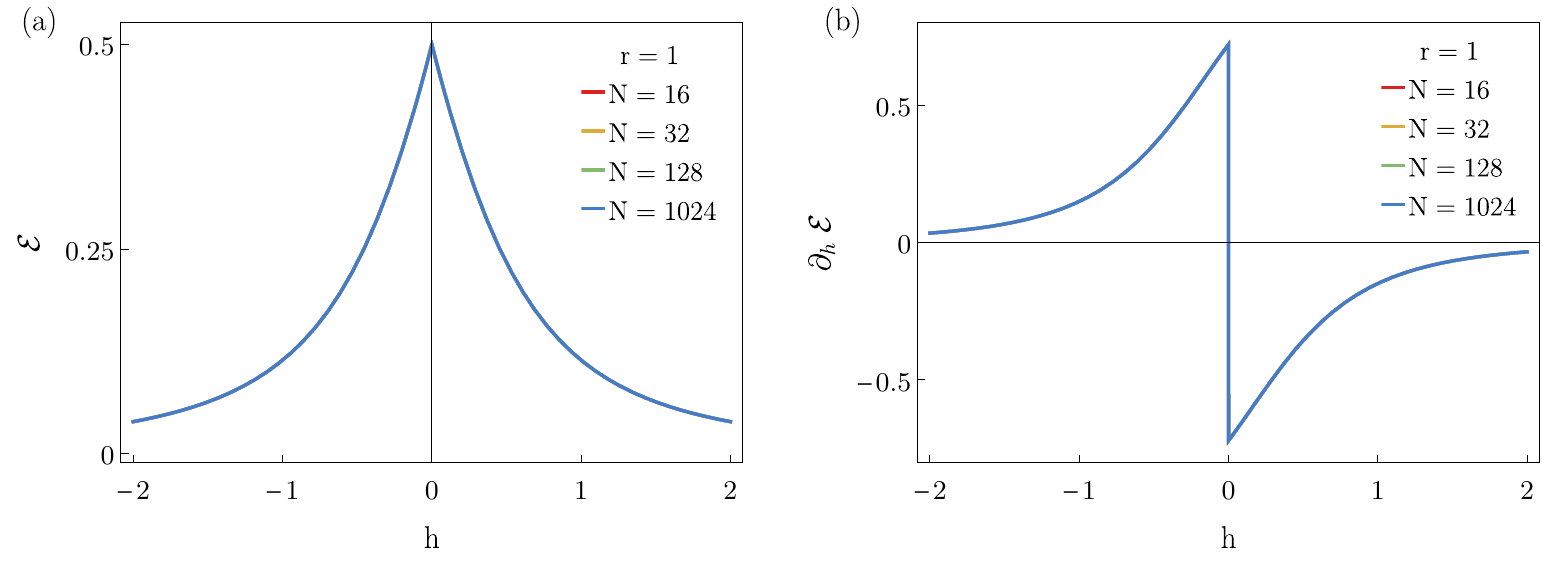}
		\caption{(a) The figure shows the entanglement per site for the halfway Ising model ($r=1$), with increasing system sizes $N=16,32,128,1024$, all of which collapse on the same line. (b) Cusp of the entanglement in (a) gives rise to a jump in the entanglement derivative.}
		\label{fig:halfwayr1Ent} 
	\end{figure*}

	As the transition in the halfway XY model is first-order, one expects that the entanglement will have a discontinuity at the transition, as it is caused by a level crossing. In this case, the ground state in the range $-\sqrt{1-r^2}\le h \le \sqrt{1-r^2}$ involves the odd sector with either one or three fermions. One could calculate the ground-state overlap with product states. But we will not proceed with that here. For $r=1$ halfway Ising model, as well as other Ising models with $n$-site interaction, the ground-state wavefunction comes from the even sector without a fermion, and for that the overlap is calculated in Sect.~\ref{sec:level2}, and hence the geometric entanglement (per site and per block of two sites) is readily available upon simple parameter optimization. As shown in Fig.~\ref{fig:halfwayr1Ent}, the entanglement develops a cusp behavior at $h=0$ and gives rise to a jump in the derivative. However, this `weak' singularity is a result that the entanglement is symmetric w.r.t. $h=0$, but it immediately decreases as soon as $h$ deviates from $0$ (i.e., with a nonzero slope). As shown in Ref.~\cite{Lahtinen2015}, at $h=0$, the state is the generalized cluster state, which exhibits the same geometric entanglement as the cluster state, and is expected to display the infinite localizable entanglement length \cite{Verstraete2004}.	Even though there is no true phase transition in the usual statistical mechanics, but there is one peculiar transition proposed by Verstraete, Martin-Delgado and Cirac \cite{Verstraete2004} in that the localizable entanglement length is infinite. This kind of transition was shown to be detectable by the geometric entanglement, displaying the weak singularity, such as the cusp \cite{Orus2010}.

	\subsection{GHZ-Cluster model}
	
	In this part, we calculate the ground-state energy of the GHZ-Cluster model, which was introduced by Wolf et al.~\cite{Wolf2006}, and examine the quantum phase transition on the phase diagram, utilizing the geometric entanglement and the energy gap. We consider a local Hamiltonian with three-site interaction constructed by the following matrix product state as its ground state:
	
	\begin{align}
	A_{0}=\Bigg(
	\begin{matrix}
	0&0\\
	1&1
	\end{matrix}\Bigg), \quad A_{1}=\Bigg(
	\begin{matrix}
	1&g\\
	0&0
	\end{matrix}\Bigg),
	\end{align}
	and the corresponding Hamiltonian possessing $\mathds{Z}_{2}$ symmetry was constructed by Wolf et al.~\cite{Wolf2006} and reads: 
	\begin{equation}
	H= \sum_{j=1}^{N} \bigg( 2(g^2 - 1) \sigma_{j-1}^{z} \sigma_{j}^{z} + (g - 1)^2 \sigma_{j-1}^{z} \sigma_{j}^{x} \sigma_{j+1}^{z} - (1 + g)^2 \sigma_{j}^{x} \bigg).
	\end{equation}
	The QPT in the model is peculiar as the ground-state energy is analytic for all range of the parameter $g$, even though the correlation length diverges at the critical point.
	
	\begin{figure*}[t]
		\includegraphics[width=1\textwidth]{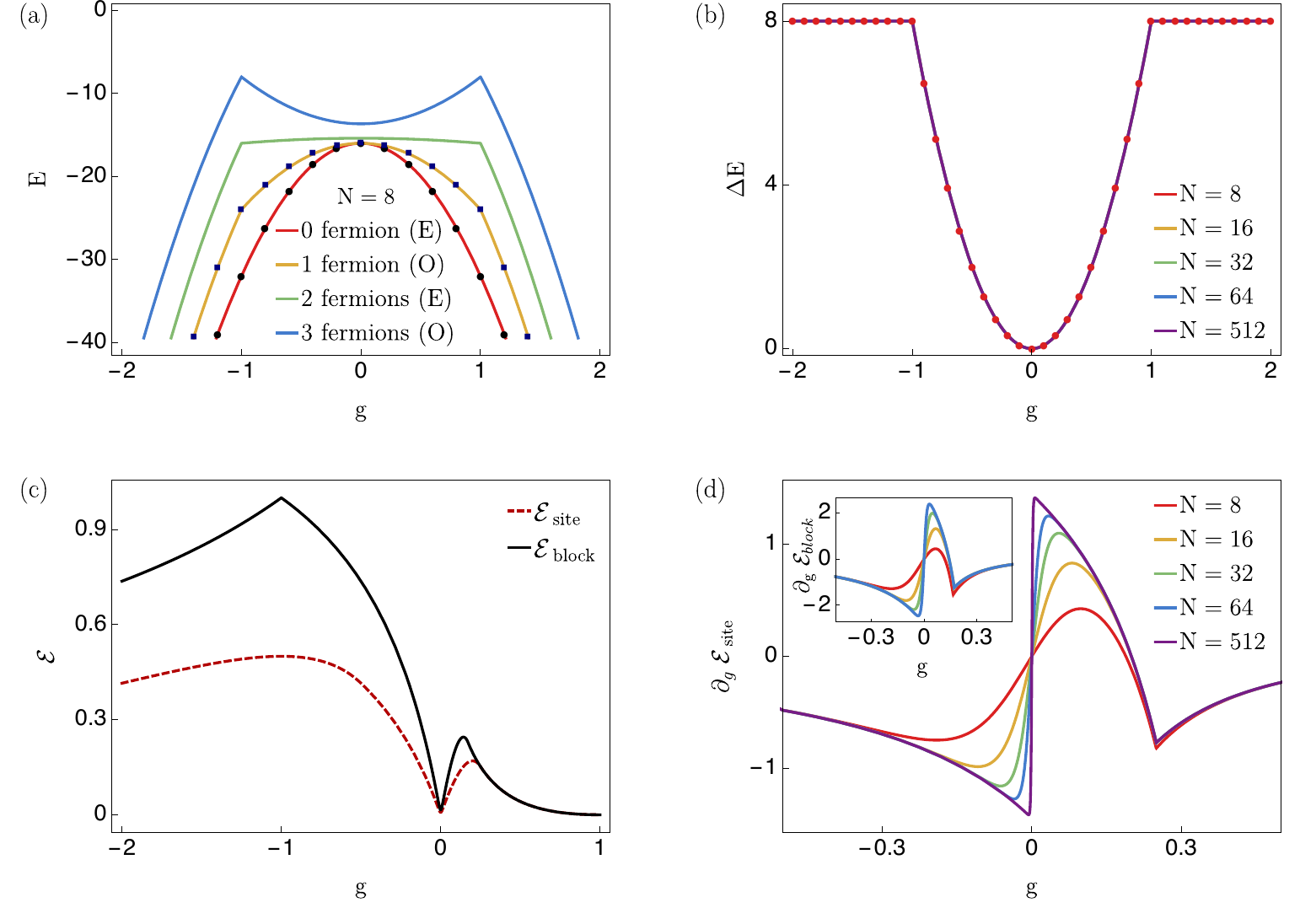}
		\caption{(a) {The lowest few energy levels for even and odd sectors in GHZ-cluster model, with $N=8$, as a function of $g$. We use E. and O. to imply Even and Odd sectors respectively.} We compare numerical exact diagonalization for lowest two energies (points) and our analytic solutions (curves). (b) The energy gap for increasing system sizes ($N=8,16,32,64,512$), interestingly, are the same. As constructed in Ref.~\cite{Verstraete2005}, the ground-state energy displays no singularity at the QPT ($g=0$). (c) The transition can be detected by the behavior of entanglement. The figure shows geometric entanglement per site (red, dashed) and per block (black, solid) for GHZ-Cluster state where $N=128$. (d) Derivative of the entanglement per site and per block (inset) close to the critical point at $g=0$, where $N=8,16,32,64,512$ is used (identified from top to bottom for $g<0$). }
		\label{fig:GHZCluster}
	\end{figure*}
	To utilize our parameterization for the model, first we rotate the Hamiltonian around the $y$ axis such that $\sigma_{x} \rightarrow \sigma_{z}$. Then we choose $N^{(x)}=2$ and a list of $J_{l}^{(x)}$, as we need two $X$ blocks and $N^{(y)}=0$ to eliminate $Y$ block. We note that one can assign the value for $h$ in terms of $g$ to generate the required Hamiltonian. Here we give the resulting parameters that give the equivalent cluster-GHZ model:
	\begin{subequations}
		\begin{align}
		h&=(1 + g)^2,\\
		N^{(x)}&=2,\
		N^{(y)}=0,\\
		J_{l}^{(x)}&=\{-2(g^2 - 1), -(g - 1)^2\},\
		J_{l'}^{(y)}=\{0\},\\
		n_{l}^{(x)}&=\{0,1\},\
		n_{l'}^{(y)}=\{0\}.
		\end{align}
	\end{subequations} 
Substituting above parameters into Eq. (\ref{eq:PXY}) yields the following Hamiltonian:
	\begin{equation}
	H= -\sum_{j=1}^{N} \bigg( -2(g^2 - 1) \sigma_{j-1}^{x} \sigma_{j}^{x} -(g - 1)^2 \sigma_{j-1}^{x} \sigma_{j}^{z} \sigma_{j+1}^{x} + (1 + g)^2 \sigma_{j}^{z} \bigg).
	\end{equation}
	{ This Hamiltonian can be diagonalized in the form of Eq.~(\ref{eq:ParHam}) with the following Bogoliubov solution, where $\varphi^{(b)}_{k}\equiv\frac{2 \pi (b+k)}{N}$,}
	\begin{align}
	\tan 2\theta_{k}^{(b)} &= -\frac{2 (g-1) \sin \varphi^{(b)}_{k} \big[(g-1) \cos \varphi^{(b)}_{k}+g+1\big]}{2 \left(g^2-1\right) \cos\varphi^{(b)}_{k}+(g-1)^2 \cos 2\varphi^{(b)}_{k}+(g+1)^2}.
	\end{align}
	The exact energy spectrum can be obtained by utilizing Eqs. (\ref{eq:PXY_betalpha}) and (\ref{eq:energyspectrum}-\ref{eq:k=N/2}). The eigenvalues in the case of even $N$ for the odd sector ($b=0$, periodic boundary conditions) and odd $N$ for the even sector ($b=1/2$, antiperiodic boundary conditions) are as follows,
	\begin{align}
	\epsilon_{k}^{(b)}= \left\{\begin{array}{lr}
	8 g^2, & \text{for } k=0 \wedge b=0\\
	8, & \text{for } k=\frac{N}{2} \wedge b=0\\
	8, & \text{for } k=\frac{N-1}{2} \wedge b=1/2
	\end{array}\right\} = 2 \alpha_{k}^{(b)},
	\end{align}
	or otherwise:
	\begin{equation}
	\epsilon_{k}^{(b)}=4 \left| 1+g^2+\left(g^2-1\right) \cos \varphi^{(b)}_{k} \right|.
	\end{equation} 
	
	The model exhibits quantum phase transition at $g_{c}=0$, and the ground state is the Greenberger-Horne-Zeilinger (GHZ) state. At $g=1$, the Hamiltonian is proportional to $\sum_j \sigma_j^z$ where all spins are in the $z$-direction; this is a paramagnetic phase. 
	At $g=-1$, the ground state is a cluster state (disordered phase), and the Hamiltonian has a $Z_2\times Z_2$ symmetry. The cluster state is a representative nontrivial $Z_2\times Z_2$ SPT state. However, the model only has $Z_2$ symmetry at $g\ne -1$. Here we also obtain the exact energy spectrum for this model using Eqs. (\ref{eq:energyspectrum}-\ref{eq:energyEven}) and analyze what ground and first excited states are composed of by examining odd/even sector and the number of fermions occupation. If we restrict ourselves to the region $-2<g<2$, we find that the ground state comes from the even sector ($b=1/2$) with no fermions and the first excited state is constructed from the odd sector ($b=0$) with one-fermion occupation. The ground state in the model has no three-fermion occupation in any finite $g$, see Fig.~\ref{fig:GHZCluster}(a). We remark that for any system size $N$ (even), the energy gap is equal to $\Delta E=8 g^2$ in the regime of $-1<g<1$; otherwise, outside that range the energy gap is always $\Delta E=8$, regardless of the system size. As already shown by construction in Ref.~\cite{Wolf2006} and confirmed here by calculation, the ground-state energy displays no singularity at the critical point $g=0$; see Fig. \ref{fig:GHZCluster}(b). It is a peculiar type of quantum phase transition, as emphasized in Ref.~\cite{Wolf2006}.

	Figure \ref{fig:GHZCluster}(c) shows the global entanglement upon using the solution which we derived in the previous section. It contains the global entanglement per site (red, dashed) and per block (black, L=2). We also examine the derivative of the entanglement~\cite{Wei2010} to study the divergence near the critical point. As shown in Fig.~\ref{fig:GHZCluster}(d), the quantum phase transition is detected at the GHZ point ($g=0$) by the behavior of entanglement. However, we note that at $g=-1$, the entanglement per block shows a cusp behavior, there is no true phase transition there. However, there is a different kind of transition there in the sense of infinite localizable entanglement length~\cite{Verstraete2004}. As remarked earlier, this kind of transition was shown to be detectable by the geometric entanglement in the form of weak singularity, such as the cusp~\cite{Orus2010}. 

	\begin{figure*}[t]
		\includegraphics[width=\textwidth]{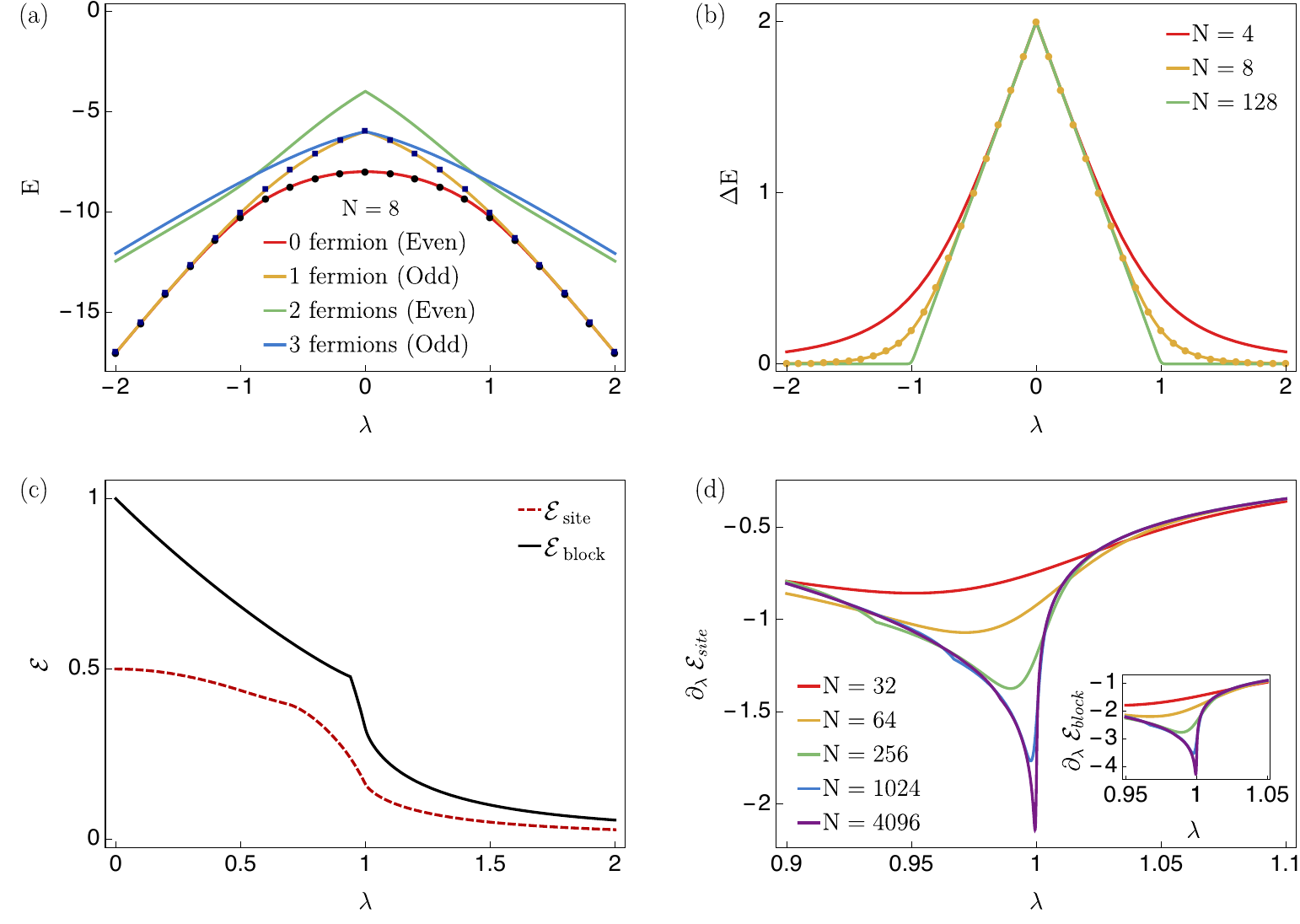}
		\caption{(a) {Lowest few energy levels vs. $\lambda$ for the SPT-AFM model with $N=8$. We show that numerical exact diagonalization for lowest two energies (points) and our analytic solutions (curves) agree. (b) The energy gap between the ground state and the first excited state as a function of $\lambda$.} The ground state is degenerate for $|\lambda| \ge 1$ in the thermodynamic limit and the energy gap becomes $\Delta E=2\big(1-\left| \lambda \right| \big) \theta(1-\left| \lambda \right|)$ where $\theta(x)=1$ if $x>0$ and zero otherwise. Thus the singularity at $\lambda=1$ signals a quantum phase transition. (c) Geometric entanglement per site (red, dashed) and per block (black, solid) for SPT-antiferromagnetic chain where $N=256$; (d) Derivative of the entanglement per site (the inset shows that for per block) where N=32, 64, 256, 1024, 4096 (from top to bottom).}
		\label{fig:SPTAFM}
	\end{figure*}
	
	\subsection{SPT-Antiferromagnetic transition}
	\label{sec:SPT} As the last example, we examine a particular quantum phase transition \cite{gu2009,pollmann2012} between a symmetry protected topological order and an antiferromagnetic phase by using the same method we derived. The specific model we study here was first discussed by Son et al.~\cite{Son2011}, who also computed the geometric entanglement per site. They showed that the transition was detected by the singular behavior of the entanglement. For completeness, we also study the spectrum and the geometric entanglement per block.
	
	In order to construct the Hamiltonian, we choose one $X$ and one $Y$ block and set $h=0$ to eliminate the transverse-field term. Parameters of the model considered are shown as follows:
	\begin{subequations}
		\label{eq:sptPar}
		\begin{align}
		h&=0,\\
		N^{(x)}&=1,\
		N^{(y)}=1,\\
		J_{l}^{(x)}&=\{1\},\
		J_{l'}^{(y)}=\{-\lambda\},\\
		n_{l}^{(x)}&=\{1\},\
		n_{l'}^{(y)}=\{0\}.
		\end{align}
	\end{subequations} 
Substituting above parameters into Eq. (\ref{eq:PXY}) yields the following Hamiltonian:
	\begin{equation}
	H= -\left( \sum_{j=1}^{N} \sigma_{j-1}^{x} \sigma^{z}_{j} \sigma_{j+1}^{x} - \lambda \sum_{j=1}^{N}\sigma_{j-1}^{y} \sigma_{j}^{y}\right). 
	\end{equation}
 This Hamiltonian can be diagonalized in the form of Eq.~(\ref{eq:ParHam}) with the following Bogoliubov solution:
	\begin{align}
	\tan 2\theta_{k}^{(b)} &= \frac{\lambda \sin \left(\frac{2 \pi (b+k)}{N}\right)+\sin \left(\frac{4 \pi (b+k)}{N}\right)}{\lambda \cos \left(\frac{2 \pi (b+k)}{N}\right)-\cos \left(\frac{4 \pi (b+k)}{N}\right)}.
	\end{align}
	The exact energy spectrum can be obtained by utilizing Eq. (\ref{eq:PXY_betalpha}) and (\ref{eq:energyspectrum}-\ref{eq:k=N/2}). The eigenvalues in the case of even $N$ for the odd sector ($b=0$, periodic boundary conditions) and odd $N$ for the even sector ($b=1/2$, antiperiodic boundary conditions) are as follows
	\begin{align}
	\epsilon_{k}^{(b)}= \left\{\begin{array}{lr}
	2 (\lambda-1), & \text{for } k=0 \wedge b=0\\
	-2 (\lambda+1), & \text{for } k=\frac{N}{2} \wedge b=0\\
	-2 (\lambda+1), & \text{for } k=\frac{N-1}{2} \wedge b=1/2
	\end{array}\right\} = 2 \alpha_{k}^{(b)},
	\end{align}
	or otherwise:
	\begin{equation}
	\epsilon_{k}^{(b)}=2 \sqrt{1+\lambda^2-2 \lambda \cos \left(\frac{6 \pi}{N}(k+b)\right)}.
	\end{equation}
	The even sector ($b=1/2$) with no fermions corresponds to the ground state energy for finite system size $N$ (even) whereas the first excited state comes from the odd sector ($b=0$) with one fermion occupation as shown in Fig. \ref{fig:SPTAFM}(a). The energy gap in this case can be obtained by calculating $\Delta E=E^{\rm lowest}_{b=0}-E^{\rm lowest}_{b=1/2}$ which is approximately $2(1-|\lambda|)$ in the region $-1/2<\lambda<1/2$ for small system size ($N$). In the thermodynamic limit ($N\rightarrow\infty$), the energy gap becomes $\Delta E=\big(1-\left| \lambda \right| \big) \big[1+\rm{sgn}(1-\left| \lambda \right|)\big]$ for all regions $-\infty<\lambda<\infty$. The critical point, $\lambda_c=1$, can be deduced from the energy gap in the thermodynamic limit; see Fig. \ref{fig:SPTAFM}(b). We also calculated geometric entanglement per site and per block, shown in Fig. \ref{fig:SPTAFM}. As can be seen in Fig. \ref{fig:SPTAFM}(d), the derivative of the entanglement per site has singularity at $\lambda=1$, at which the quantum phase transition occurs between the cluster and the antiferromagnetic phases. We note that as the antiferromagnetic phase is involved in the model, in order to compute entanglement per site, we use the closest product state of the form $\ket{\Phi}=\prod_{i}\ket{\phi^{[2i-1,2i]}}$ with $\ket{\phi^{[2i-1,2i]}}=(\alpha \ket{\uparrow}+\beta \ket{\downarrow})(\gamma\ket{\uparrow}+\delta\ket{ \downarrow})$. The entanglement derivative w.r.t. $\lambda$ clearly also shows the development of divergence at $\lambda=1$ as the system size $N$ increases. The representative state in the SPT phase is the 1D cluster state~\cite{raussendorf2001,else2012}, which we also have seen in previous subsection. We remark that there is a weak singularity in the entanglement per block around $\lambda\approx 0.94$, but we cannot identity the state there and do not know the nature of this singularity. It might be a transition in localizable entanglement, but that requires further investigation.
		\begin{figure}[t]
		\centering 
		\includegraphics[width=\textwidth]{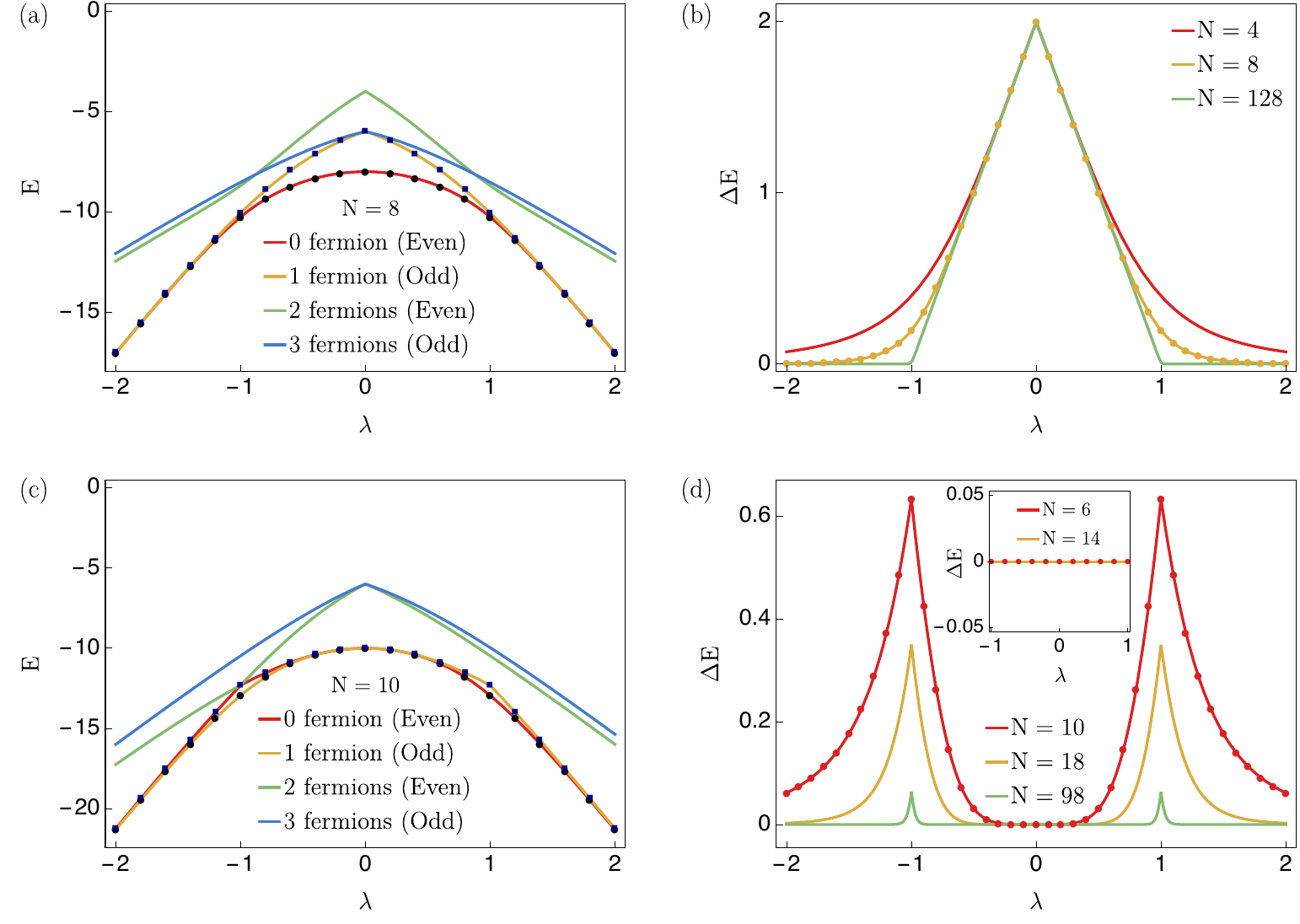}
		\caption{\label{fig:halfwaySPTgap} The lowest two levels for even and odd sectors with (a): $N=8$; (c): $N=10$, for the SPT-antiferromagnetic chain with halfway interaction. Similar to SPT-AFM model the ground state comes from the even sector ($b=1/2$) with no fermion for $N=4m$ (with $m=1,2...$). In the case of $N=10$, ($N=2(4m+1)$), the lowest zero-fermion and one-fermion states become degenerate except in the vicinity of $\lambda=1$. Interestingly, at the point $\lambda=-1$ ground state energy is constructed by the odd sector with one-fermion occupation, whereas at $\lambda=1$, the ground state energy comes from the even sector with zero fermion occupation. We confirm our analytic solutions (curves) with the results obtained from numerical exact diagonalization for lowest two energies and energy gap (points). We note that the energy gap has different characteristics for (b) $N=4m$ and (d) $N=2(4m+1)$ and the inset figure illustrates $N=2(4m-1)$ case. The latter is gapless for all range of $\lambda$. In the case of $N=4m$, the ground state is degenerate for $|\lambda| \ge 1$. With increasing system size energy gap closes continuously. Thus the singularity at $\lambda_c=\pm1$ signals a quantum phase transition.}
	\end{figure}

	\subsection{Halfway antiferromagnetic-SPT model}
	\label{sec:halfwaySPT} Beyond reproducing results by Son et al.~\cite{Son2011}, we also examine a slight variation in the model, where, instead of $XZX$, the \textit{halfway interaction} for $X$ blocks is considered: 
	\begin{equation}
	H= -\left(\sum_{j=1}^{N} \sigma_{j-1}^{x} \sigma^{z}_{j} \ldots \sigma^{z}_{j+(N/2)-2} \sigma_{j+(N/2)-1}^{x} - \lambda \sum_{j=1}^{N}\sigma_{j-1}^{y} \sigma_{j}^{y}\right).
	\end{equation}
	The parameters for this model can be defined as follows:
	\begin{subequations}
		\begin{align}
		h&=0,\\
		N^{(x)}&=1,\
		N^{(y)}=1,\\
		J_{l}^{(x)}&=\{1\},\
		J_{l'}^{(y)}=\{-\lambda\},\\
		n_{l}^{(x)}&=\{N/2-1\},\
		n_{l'}^{(y)}=\{0\}.
		\end{align}
	\end{subequations}
	The model can be exactly diagonalized with the following Bogoliubov solution:
	\begin{align}
	\tan 2\theta_{k}^{(b)} &= \frac{\lambda \sin \left(\frac{2 \pi (b+k)}{N}\right)+\sin \left(\pi (b+k)\right)}{\lambda \cos \left(\frac{2 \pi (b+k)}{N}\right)-\cos \left(\pi (b+k)\right)}.
	\end{align}
	{The exact energy spectrum can be obtained by utilizing Eqs. (\ref{eq:PXY_betalpha}) and (\ref{eq:energyspectrum}-\ref{eq:k=N/2}). The eigenvalues in the case of even $N$ for the odd sector ($b=0$, periodic boundary conditions) and odd $N$ for the even sector ($b=1/2$, antiperiodic boundary conditions) are as follows
	\begin{align}
	\epsilon_{k}^{(b)}= \left\{\begin{array}{lr}
	2 (\lambda-1), & \text{for } k=0 \wedge b=0\\
	-2 \left[\lambda+(-1)^{N/2}\right], & \text{for } k=\frac{N}{2} \wedge b=0\\
	-2 \lambda, & \text{for } k=\frac{N-1}{2} \wedge b=1/2
	\end{array}\right\} = 2 \alpha_{k}^{(b)},
	\end{align}
	or otherwise:
	\begin{equation}
	\epsilon_{k}^{(b)}=2 \sqrt{1+\lambda^2-2 \lambda \cos \left(\frac{(2+N) \pi}{N}(k+b)\right)}.
	\end{equation}
		\begin{figure}[t]
		\centering
		\includegraphics[width=0.5\textwidth]{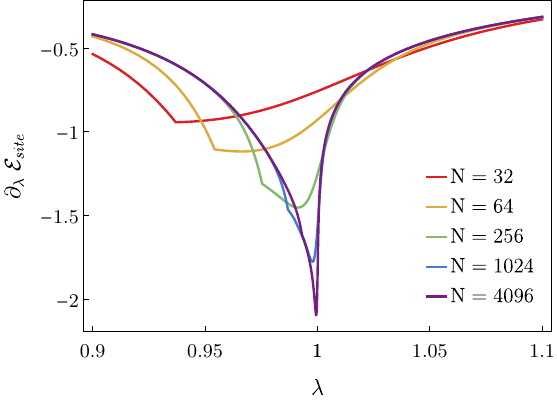}
		\caption{\label{fig:halfwaySPT} Derivative of the entanglement per site where N=32, 64, 256, 1024, 4096 (from top to bottom) for SPT-antiferromagnetic chain with halfway interaction.}
	\end{figure}
Similar to SPT-AFM model, the ground state is constructed from the even sector ($b=1/2$) with no fermions as the first excited state comes from the odd sector ($b=0$) with one-fermion occupation for $N=8$, see Fig. \ref{fig:halfwaySPTgap}(a). On the other hand, in the case of $N=10$, the lowest zero-fermion and one-fermion states become degenerate except in the vicinity of $\lambda=1$. Interestingly, at the point $\lambda=-1$ ground state energy is constructed by the odd sector with one-fermion occupation, whereas at $\lambda=1$, the ground state energy comes from the even sector with zero fermion occupation, see Fig. \ref{fig:halfwaySPTgap}(c). This model does not exhibit the peculiarity discussed in Eq. (\ref{eq:threeFermions}), where the odd sector has three-fermion occupation as the lowest-energy state. We note that the energy gap has different characteristics depending on the system sizes (even): $N=4m$, $N=2(4m+1)$, and $N=2(4m-1)$ (with $m=1,2...$). The latter is gapless for all range of $\lambda$, whereas the case of $N=2(4m+1)$ displays a peak at the $\lambda=1$, as shown in Fig.~\ref{fig:halfwaySPTgap}(d). With increasing system sizes, the peak approaches to zero, and in the thermodynamic limit, both cases become gapless. However, the case of $N=4m$ exhibits similar behavior to the SPT-AFM model with critical points $\lambda_c=\pm1$, see Fig.~\ref{fig:halfwaySPTgap}(b). The ground state is degenerate for $|\lambda| \ge 1$ in the thermodynamic limit and the energy gap becomes $\Delta E=2\big(1-\left| \lambda \right| \big) \theta(1-\left| \lambda \right|)$, where $\theta(x)=1$ if $x>0$ and zero otherwise. Thus, the singularity at $\lambda=1$ signals a quantum phase transition. This is in contrast to the halfway XY-model, discussed in Sect.~\ref{sec:XYhalfway}, that the halfway interaction prevents the model from undergoing a quantum phase transition but rather helps to exhibit a first-order transition across the Barouch-McCoy circle. The quantum phase transition (for $N=4m$ case) in the halfway SPT-AFM model can be confirmed by the behavior of entanglement as well. With an increasing system size, the derivative of the entanglement per site develops a singularity at $\lambda_c=1$, at which the quantum phase transition takes place; see Fig. \ref{fig:halfwaySPT}.}
	
	\section{Conclusion}
	
	In this work, we introduced a convenient parameterization for a general class of exactly solvable spin chains, which we called the cluster-XY models. We reviewed the procedure to diagonalize these spin chains and obtained the energy spectrum, the ground-state energy, the ground-state wavefunctions, and the energy gap. We illustrated the subtlety in determining the true ground state, as it can come from two different sectors, with different numbers of fermions. The quantum phase transitions can be studied from the energy gap in the thermodynamic limit. Furthermore, we employed the geometric measure of entanglement per site/block for quantifying entanglement in the many-body system. We presented detailed calculations for the overlap of the ground states with two different types of product states. Using these, we examined the global entanglement near the quantum critical point in several illustrative models that include the three-site interacting XY model, the XY model with halfway interaction, the GHZ-cluster model, and the SPT-AFM models (and a variation in the last model).
	
	Among the above models, the XzY model possesses a $Z_2\times Z_2$ symmetry and exhibits transitions from nontrivial SPT phase to a trivial paramagnetic phase. Such a transition is expected to exist in all other finite-range Xz...zY models. However, it does not appear in the halfway XY model. Instead, the halfway XY model exhibits a first-order transition across the Barouch-McCoy circle, on which it was only a crossover in the standard XY model. However, the halfway Ising model has no such transition.
	
	The GHZ-cluster model was constructed in Ref.~\cite{Verstraete2005} to exhibit a QPT but without singularity in ground-state energy. Geometric entanglement was able to detect such QPT~\cite{Wei2010}. The SPT-AFM model is an interesting example that has a transition between a symmetry-protected topological phase and a symmetry-breaking phase~\cite{Son2011}. We not only reproduced the entanglement per site but also presented results using the entanglement per block {and examined the spectrum and the energy gap}. Both quantities display singularity near the critical point. Furthermore, we also studied a peculiar variation, where the cluster interaction $XZX$ is replaced by a halfway interaction. In contrast to the halfway Ising model, this halfway SPT-AFM model exhibits a QPT. Our study on arbitrary $n$-site XY model generalizes previous study on the XY model via the geometric entanglement~\cite{Wei2005}. These examples we gave demonstrate the usefulness of our general results on entanglement in the family of the generalized XY-cluster models.
	
	Regarding the entanglement per block, we were able to obtain analytic results for a block of two sites. The two-site state can be generally entangled, but can also be set to be a product state. The latter is useful for the geometric entanglement per site in the case of antiferromagnetic ground states, as the globally the closest product state cannot be translationally invariant. Even though numerically one can compute per block of any number of sites, it would be interesting to derive analytically the overlap with block product state composed of any number of sites in a block. Then, the entanglement under RG can be studied in further detail. We leave it for future exploration. 
	
\begin{acknowledgements}
This work was supported by National Science Foundation via Grants No. PHY 1620252 and No. PHY 1314748. T.-C.W. also acknowledges support of a SUNY seed grant. A.D. hereby acknowledges Fulbright scholarship, granted by the US Department of State's Bureau of Educational and Cultural Affairs, to fund the author during this research in Stony Brook University. Some results are based on his master's thesis submitted to the Graduate School at Stony Brook University in 2016 \cite{deger2016}.
\end{acknowledgements}

\end{document}